\documentclass[journal]{IEEEtran}
\ifCLASSINFOpdf
\else
\fi
\usepackage{amssymb}
\usepackage{amsmath}
\usepackage{amsthm}

\usepackage{cite}

\usepackage{hyperref}
\usepackage{multirow}
\usepackage{booktabs}

\usepackage{tikz}
\usetikzlibrary{plotmarks}
\usetikzlibrary{calc}
\usetikzlibrary{arrows}
\usetikzlibrary{arrows,intersections}
\usetikzlibrary{arrows.meta}
\usetikzlibrary{shapes,decorations}
\usetikzlibrary{positioning}
\usepackage{pgfplots}
\tikzset{sin v source/.style={
  circle,
  draw,
  append after command={
    \pgfextra{
    \draw
      ($(\tikzlastnode.center)!0.5!(\tikzlastnode.west)$)
       arc[start angle=180,end angle=0,radius=0.425ex]
      (\tikzlastnode.center)
       arc[start angle=180,end angle=360,radius=0.425ex]
      ($(\tikzlastnode.center)!0.5!(\tikzlastnode.east)$)
    ;
    }
  },
  scale=1.5,
 }
}

\pgfplotsset{compat=1.14}
\usepackage{circuitikz}
\usepackage{tikz}

\begin{document}
%

\title{Convex Relaxations of Probabilistic AC Optimal Power Flow for Interconnected \\ AC and HVDC Grids}

%
%

\author{Andreas Venzke,~\IEEEmembership{Student Member, IEEE,}
	and Spyros Chatzivasileiadis,~\IEEEmembership{Senior Member, IEEE}
	\thanks{A. Venzke and S. Chatzivasileiadis are with the Center for Electric Power and Energy, Technical University of Denmark, Kgs. Lyngby, Denmark. This work is supported by the multiDC project, funded by Innovation Fund Denmark, Grant Agreement No. 6154-00020B.}
	}

%
%

\markboth{}%
{}
%



\maketitle


\begin{abstract}
High Voltage Direct Current (HVDC) systems interconnect AC grids to increase reliability, connect offshore wind generation, and enable coupling of electricity markets. Considering the growing uncertainty in power infeed and the complexity introduced by additional controls, robust decision support tools are necessary. This paper proposes a chance constrained AC-OPF for AC and HVDC grids, which considers wind uncertainty, fully utilizes HVDC control capabilities, and uses the semidefinite relaxation of the AC-OPF. We consider a joint chance constraint for both AC and HVDC systems, we introduce a piecewise affine approximation to achieve tractability of the chance constraint, and we allow corrective control policies for HVDC converters and generators to be determined. An active loss penalty term in the objective function and a systematic procedure to choose the penalty weights allow us to obtain AC-feasible solutions. We introduce Benders decomposition to maintain scalability. Using realistic forecast data, we demonstrate our approach on a 53-bus and a 214-bus AC-DC system, obtaining tight near-global optimality guarantees. With a Monte Carlo analysis, we show that a chance constrained DC-OPF leads to violations, whereas our proposed approach complies with the joint chance constraint.
\end{abstract}


%
\IEEEpeerreviewmaketitle

\section{Introduction}
\IEEEPARstart{T}he increase of uncertain renewable generation and the growing electricity demand lead power systems to operate closer to their limits \cite{brown2016optimising}. To maintain a secure operation, significant investment in new transmission capacity and an improved utilization of existing assets are necessary. The High Voltage Direct Current (HVDC) technology is a promising candidate for enabling increased penetration of volatile renewable energy sources and providing controllability in power system operation. In China, in order to transport large amounts of power, e.g. wind, from geographically remote areas to load centers, significant HVDC transmission capacity has been built \cite{bogdanov2016north}. A European HVDC grid is envisioned, extending several of the point-to-point connections already in operation to a multi-terminal grid \cite{pierri2017challenges}. In this work, we address the challenge of the operation of such a system under uncertainty. To this end, we propose a tractable formulation of the chance constrained AC optimal power flow (OPF) for interconnected AC and HVDC grids which includes HVDC corrective control capabilities.

 Several works in the literature integrate models of HVDC grids in the AC-OPF formulation \cite{beerten2012generalized, chatzivasileiadis2012sepope, chatzivasileiadis2014security}. The work in \cite{beerten2012generalized} introduces a generalized steady-state Voltage Source Converter (VSC) multi-terminal DC model which can be used for sequential AC/DC power flow algorithms.  The work in \cite{chatzivasileiadis2012sepope} presents probably the first formulation of a security-constrained AC-OPF which considers the corrective control capabilities of the HVDC converters. This is further extended in \cite{chatzivasileiadis2014security}, proposing linear approximations. 
 
 To account for uncertainty, chance constraints can be included in the OPF, defining a maximum allowable probability of constraint violation. Using the DC-OPF approximation, the work in \cite{bienstock2014chance} formulates a chance-constrained DC-OPF assuming Gaussian distribution of renewable forecast errors and the work in \cite{lubin2015robust} proposes a robust DC-OPF.  In literature, the application of chance constraints in the context of interconnected AC and HVDC grids is limited to the DC-OPF formulation \cite{vrakopoulou2013HVDC, wiget2014probabilistic}. Ref.~\cite{vrakopoulou2013HVDC} proposes a tractable formulation of a probabilistic security constrained DC-OPF with HVDC lines. This framework is extended to include HVDC grids in Ref.~\cite{wiget2014probabilistic}. There are two main motivations to use an AC-OPF formulation. First, the DC power flow formulation neglects voltage magnitudes, reactive power and system losses and can lead to substantial errors \cite{dvijotham2016}. Second, the AC-OPF formulation allows to represent and utilize the voltage and reactive power control capabilities of the HVDC converters. The works \cite{RoaldCCACOPF,Lorca} address the chance constrained AC-OPF problem for AC grids. Using a linearization of the AC system state around the operating point, the work in \cite{RoaldCCACOPF} achieves a tractable formulation of the chance constraints assuming Gaussian distribution of the uncertainty. The work in \cite{Lorca} proposes a two-stage adaptive robust optimization model for the multi-period AC-OPF using semidefinite and second-order cone relaxations and a budget uncertainty set. The scope of this work is to propose a tractable formulation for the chance constrained AC-OPF for interconnected AC and HVDC grids.

Several works \cite{baradar2013second,Moghadasi,bahrami2017semidefinite} use convex optimization techniques for the AC-OPF problem for AC and HVDC grids without considering uncertainty. A convex formulation can provide guarantees for global optimality. The work in \cite{baradar2013second} proposes a second-order cone relaxation. In \cite{Moghadasi}, a semidefinite formulation for the voltage-stability constrained OPF is proposed. The work in \cite{bahrami2017semidefinite} introduces a convex relaxation of the AC-OPF problem for interconnected AC and HVDC grids, using the semidefinite relaxation technique in \cite{lavaei2012zero} and including a detailed HVDC converter model. The work in \cite{Bahrami2018} extends this framework towards a security constrained unit commitment problem under uncertainty. However, it is assumed that the forecast errors and resulting generation and load mismatch are balanced at each bus internally via curtailment, energy storage or reserve units located at this bus. Hence, the line flows and voltages do not change as a result of different uncertainty realizations. This assumption is restrictive and can lead to high levels of curtailment in practice, e.g. for offshore wind generation without energy storage.

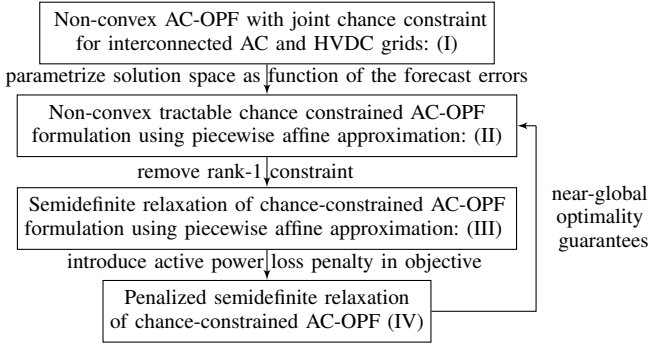
\begin{figure}
    \begin{footnotesize}
    \tikzstyle{block} = [draw, fill=white, rectangle, 
    minimum height=0em, minimum width=0]
\tikzstyle{sum} = [draw, fill=blue!20, circle, node distance=4cm]
\tikzstyle{input} = [coordinate]
\tikzstyle{output} = [coordinate]
\tikzstyle{pinstyle} = [pin edge={to-,thin,black}]

\begin{tikzpicture}[node distance=0em,>=latex']
\node [block] (ACOPF) {$\begin{matrix} \text{ Non-convex AC-OPF with joint chance constraint} \\ \text{ for interconnected AC and HVDC grids: (I) } \end{matrix}$};
\node [block, below = 0.4cm of ACOPF] (CCACOPF) {$\begin{matrix} \text{ Non-convex tractable chance  constrained AC-OPF } \\ \text{ formulation using piecewise affine approximation: (II) } \end{matrix}$};
\node [block, below = 0.4cm of CCACOPF] (CCACOPFAP) {$\begin{matrix} \text{Semidefinite relaxation of chance-constrained AC-OPF} \\ \text{formulation using piecewise affine approximation: (III) }  \end{matrix}$};
\node [block, below = 0.4cm of CCACOPFAP] (CCACOPFAP2) {$\begin{matrix} \text{ Penalized semidefinite relaxation } \\ \text{of chance-constrained AC-OPF (IV)}  \end{matrix}$};
\draw [->] (ACOPF) -- (CCACOPF);
\draw [->] (CCACOPF) -- (CCACOPFAP);
\draw [->] (CCACOPFAP) --(CCACOPFAP2);
\node [above = 0.001 cm of CCACOPF] (sssssss) {  parametrize solution space as function of the forecast errors};
\node [above = 0.02 cm of CCACOPFAP] (sssssss) { \, \, remove rank-1 constraint \quad \, \, \, \, };
\node [above = 0.001 cm of CCACOPFAP2] (sssssss) { \hspace{0.1cm}  introduce active power loss penalty in objective};
\node [input,right = 0.25 cm of CCACOPF] (tmp) {};
\node [right = 0.35 cm of CCACOPFAP] (tmp2) {$\begin{matrix} \text{near-global} \\ \text{optimality} \\ \text{ guarantees} \end{matrix}$};
\draw (CCACOPFAP2) -| (tmp);
\draw [->] (tmp) -- (CCACOPF);
\end{tikzpicture}
    \end{footnotesize}
    \vspace{-0.65cm}
    \caption{Using a piecewise affine approximation, a tractable formulation of the chance constrained AC-OPF for interconnected AC and HVDC grids is proposed. This problem is relaxed by dropping the non-convex rank constraint. If the obtained matrices are not rank-1, a penalized semidefinite relaxation is proposed that can recover rank-1 solutions and upper bounds the sub-optimality with respect to (II).}
    \label{Relaxation}
\end{figure}
An overview of the proposed methodology to achieve a tractable formulation of the chance constrained AC-OPF for interconnected AC and HVDC grids is illustrated in Fig.\ref{Relaxation}. First, to address uncertainty in wind power injections, we include a joint chance constraint which ensures that the AC and HVDC system constraints are satisfied for a defined probability. We use a scenario-based rectangular uncertainty set. As the AC power flow is non-linear, a suitable approximation of the system state as a function of the uncertain variables has to be introduced. The work in \cite{RoaldCCACOPF} proposes a linearization around the forecasted operating point. In this work, to accurately model large uncertainty deviations, we use a piecewise linear approximation between the forecasted operating point and the vertices of a rectangular or polyhedral uncertainty set. Then, we relax the non-convex chance constrained AC-OPF formulation to a semidefinite program. 
As the resulting chance constraints are convex, we enforce them only for the vertices of the uncertainty set \cite{margellos2014road}. For the semidefinite relaxation of the AC-OPF to be feasible to the non-convex AC-OPF, the rank of the introduced matrix variables has to be equal to 1 \cite{lavaei2012zero}. 
In case we do not obtain rank-1 solution matrices, we propose a systematic method using a penalty term on active power losses to recover rank-1 solution matrices. 
As we solve a convex relaxation, we can derive near-global optimality guarantees that upper bound the distance to the global optimum of the non-convex AC-OPF. We will show in our simulation studies that the obtained set-points from the piecewise affine approximation lead to AC power flow solutions which respect the joint chance constraint violation probability.

In our previous work \cite{venzke2017}, we introduced a comprehensive framework to handle chance constraints for the semidefinite relaxation of the AC-OPF, including Gaussian distributions, while in \cite{Halilbasic2018} we addressed issues related to the affine approximations of the chance-constrained OPF for the second-order cone formulation. In \cite{venzke2018}, we further extended the work of \cite{venzke2017} by considering security constraints. In this paper, we introduce a tractable formulation of the AC-OPF under uncertainty for interconnected AC and HVDC grids. The main contributions of our work are:
\begin{itemize}
\item To the authors' knowledge, this is the first paper that proposes a tractable formulation of the chance constrained  AC-OPF for interconnected AC and HVDC grids.
\item We introduce the decomposition of the positive semidefinite matrix variables in separate submatrices, each corresponding to an individual AC or HVDC grid. This technique increases scalability and improves numerical stability.
\item We also introduce piecewise affine corrective control policies for active and reactive power of HVDC converters in addition to generator active power and voltage, and utilize wind farm reactive power capabilities, to react to wind forecast errors. 
\item We enable parallel computation through Benders decomposition to address high-dimensional uncertainty. To this end, we formulate one subproblem for each vertex of the rectangular uncertainty set and define suitable feasibility and optimality cuts. We  apply the decomposition strategy on a 214-bus AC-DC system.
\item We propose a systematic method to identify suitable penalty weights to obtain rank-1 solution matrices, by introducing a penalty term on active power losses. We show that this penalty term obtains significantly tighter near-global optimality guarantees than a reactive power penalty proposed in literature.
\item Using realistic day-ahead forecast data, we demonstrate the performance of our approach on two 24 bus systems interconnected with an HVDC grid and offshore wind generation. With a Monte Carlo analysis and using AC-DC power flows of MATACDC \cite{beerten2015development}, we compare our approach to a chance constrained DC-OPF formulation. We find that our approach complies with the considered joint chance constraint whereas the DC-OPF leads to violations. For the considered time steps, the obtained near-global optimality guarantees are higher than 99.5\%.
\item To match the empirical closely with the maximum allowable joint chance constraint violation probability, we propose a heuristic adjustment procedure for the scenario-based uncertainty set by discarding worst-case samples. This allows us to reduce the cost of uncertainty.
\end{itemize}

The rest of this paper is organized as follows. Section~\ref{II} states the semidefinite relaxation of the AC-OPF for interconnected AC and HVDC grids and includes the joint chance constraint. Section~\ref{III} defines the scenario-based uncertainty set, the piecewise affine approximation and formulates corrective control policies. To achieve tractability, a method from randomized and robust optimization is applied. Then, Benders decomposition is applied to the AC-OPF formulation. Section~\ref{IV} presents the results. Section~\ref{V} concludes.
\section{AC Optimal Power Flow Formulation} \label{II}
In this section, we formulate the semidefinite relaxation of the AC-OPF for interconnected AC and HVDC grids and include a joint chance constraint. Ref.~\cite{lavaei2012zero} proposes a semidefinite relaxation of the AC-OPF by formulating the OPF as a function of a positive semidefinite matrix variable $W$ describing the product of real and imaginary bus voltages. The convex relaxation is obtained by dropping the rank-1 constraint on the matrix $W$. We build our formulation upon \cite{bahrami2017semidefinite}, which extends the initial work \cite{lavaei2012zero} to AC-DC grids. Among the contributions of this paper is that we decompose the problem, using one matrix $W^i$ for each AC and HVDC grid instead of one matrix $W$ for the entire system, as in \cite{bahrami2017semidefinite}, and we include a chordal decomposition of the semidefinite constraints on matrices $W^i$. This allows for scalability and increased numerical stability. 
\subsection{Semidefinite Relaxation of AC Optimal Power Flow for Interconnected AC and HVDC Grids}  \label{II.1}
The system of interconnected AC and HVDC grids consists of a number of $n_{\text{grid}}$ AC and HVDC grids which are interfaced by a number of $n_c$ HVDC  converters. Each HVDC grid is modeled similar to an AC grid, but with purely resistive transmission lines, and generators operating at unity power factor. Each AC and HVDC grid $i$ is comprised of $\mathcal{N}^{i}$ buses and $\mathcal{L}^{i}$ lines. The set of buses with a generator connected is denoted with $\mathcal{G}^{i}$. The following auxiliary variables are introduced for each bus $k \in \mathcal{N}^{i}$ and line $(l,m) \in \mathcal{L}^{i}$:
\begin{align*}
	Y_k^{i} &:= e_k e_k^T Y^{i}, \, \, \, 	Y_{lm}^{i} := (\bar{y}_{lm}^{i} + y_{lm}^{i}) e_l e_l^T - (y_{lm}^{i}) e_l e_m^T    \\
	\textbf{Y}_k^{i} & := \dfrac{1}{2} \begin{bmatrix} \Re \{Y_k^{i} + (Y_k^{i})^{T}\} & \Im \{ (Y_k^{i})^{T} - Y_k^{i} \} \\ \Im \{ Y_k^{i} - (Y_k^{i})^{T}\} & \Re \{Y_k^{i} + (Y_k^{i})^{T}\} \end{bmatrix} \\
	\textbf{Y}_{lm}^{i} &:= \dfrac{1}{2} \begin{bmatrix} \Re \{Y_{lm}^{i} + (Y_{lm}^{i})^{T}\} & \Im \{ (Y_{lm}^{i})^{T}- Y_{lm}^{i} \} \\ \Im \{ Y_{lm}^{i} - (Y_{lm}^{i})^{T}\} & \Re \{Y_{lm}^{i} + (Y_{lm}^{i})^{T}\} \end{bmatrix} \\
	\bar{\textbf{Y}}_k^{i} &:= \dfrac{-1}{2} \begin{bmatrix} \Im \{Y_k^{i} + (Y_k^{i})^T\} & \Re \{ Y_k^{i} - (Y_k^{i})^T \} \\ \Re \{ (Y_k^{i})^T - Y_k^{i}\} & \Im \{Y_k^{i} + (Y_k^{i})^T\} \end{bmatrix} \\
	\bar{\textbf{Y}}_{lm}^{i} &:= \dfrac{-1}{2} \begin{bmatrix} \Im \{Y_{lm}^{i} + (Y_{lm}^{i})^{T}\} & \Re \{ Y_{lm}^{i} - (Y_{lm}^{i})^{T}  \} \\ \Re \{ (Y_{lm}^{i})^{T} - Y_{lm}^{i} \} & \Im \{Y_{lm}^{i} + (Y_{lm}^{i})^{T}\} \end{bmatrix} \\
	\textbf{M}_k &:= \begin{bmatrix} e_k e_k^T & 0 \\ 0 & e_k e_k^T \end{bmatrix} \\
	\textbf{M}_{lm} &:= \begin{bmatrix} (e_l - e_m) (e_l - e_m)^T & 0 \\ 0 & (e_l - e_m) (e_l - e_m)^T \end{bmatrix} 
\end{align*}
For each AC and HVDC grid $i$, matrix $Y^{i}$ denotes the bus admittance matrix, $e_k$ the k-th basis vector, $\bar{y}_{lm}^{i}$ the shunt admittance of line $(l,m) \in \mathcal{L}^{i}$ and $y_{lm}^{i}$ the series admittance. The non-linear AC-OPF problem for interconnected AC and HVDC grids can be written as
\begin{align}
	\min_{W^{i}} \sum_{i=0}^{n_{\text{grid}}} \sum_{k \in \mathcal{G}^{i}} \{ c_{k2}^{i} (\text{Tr} \{ \textbf{Y}_k^{i} W^{i}\} + P^{i}_{D_k})^2 + \hphantom{{}=Q_{\text{min},k} } \notag \\[-3\jot]
	c_{k1}^{i}  (\text{Tr} \{ \textbf{Y}_k^{i} W^{i}\} + P_{D_k}^{i}) + c_{k0}^{i} \} \label{MinGen}
\end{align}
subject to the following constraints for each bus $k \in \mathcal{N}^{i}$ and line $(l,m) \in \mathcal{L}^{i}$ of each power grid $i$:
\begin{align}
	\underline{P}^{i}_{G_k}  \leq \text{Tr} \{ \textbf{Y}^{i}_k W^{i}\} + P_{D_k}^{i} \leq \overline{P}^{i}_{G_k}  \label{PBal} \\
	\underline{Q}^{i}_{G_k}   \leq \text{Tr} \{ \bar{\textbf{Y}}_k^{i} W^{i}\} + Q^{i}_{D_k} \leq \overline{Q}^{i}_{G_k} \label{QBal}      	\end{align}
	\begin{align}
	(\underline{V}_k^{i})^2 \leq \text{Tr} \{ \textbf{M}_k^{i} W^{i}\} \leq (\overline{V}_k^{i})^2 \label{VCon}                                                 \\
	-\overline{P}_{lm}^{i} \leq \text{Tr} \{ \textbf{Y}^{i}_{lm} W^{i}\} \leq \overline{P}_{lm}^{i}  \label{PlmCon}                                \\
	\text{Tr} \{ \textbf{Y}_{lm}^{i} W^{i}\}^2 +   \text{Tr} \{ \bar{\textbf{Y}}_{lm}^{i} W^{i}\}^2 \leq (\overline{S}_{lm}^{i})^2 \label{SlmCon} \\
	W^{i} = [\Re \{ \textbf{V}^{i} \} \, \Im \{ \textbf{V}^{i} \}] \, \, [\Re \{ \textbf{V}^{i} \} \, \Im \{ \textbf{V}^{i} \}]^T  \label{WTT}
\end{align} 
The objective \eqref{MinGen} minimizes generation cost, where $c^{i}_{k2}$, $c^{i}_{k1}$ and $c^{i}_{k0}$ are quadratic, linear and constant cost variables associated with power production of generator $k \in \mathcal{G}^{i}$. The terms $P_{D_k}^{i}$ and $Q_{D_k}^{i}$ denote the active and reactive power consumption at bus $k \in \mathcal{N}^i$. Constraints \eqref{PBal} and \eqref{QBal} include the nodal active and reactive power flow balance; $\underline{P}^{i}_{G_k}$, $\overline{P}^{i}_{G_k}$, $\underline{Q}^{i}_{G_k}$ and  $\overline{Q}^{i}_{G_k}$ are generator limits for minimum and maximum active and reactive power, respectively. The bus voltages are constrained by \eqref{VCon} with corresponding lower and upper limits $\underline{V}^{i}_k$, $\overline{V}^{i}_k$. The active and apparent power branch flow $P_{lm}^{i}$ and $S^{i}_{lm}$ on line $(l,m) \in \mathcal{L}^{i}$ are limited by $\overline{P}_{lm}^{i}$ \eqref{PlmCon} and $\overline{S}^{i}_{lm}$ \eqref{SlmCon}, respectively. The vector of complex bus voltages is denoted with $\textbf{V}^{i}$. To obtain an optimization problem linear in $W^{i}$, the objective function is reformulated using Schur's complement introducing auxiliary variables $\alpha^{i}$ for each power grid $i$:
\begin{align}
	\min_{\alpha^{i}, W^{i}} \sum_{i=0}^{n_{\text{grid}}} \sum_{k \in \mathcal{G}^{i}} \alpha_k^{i} \hphantom{100000000000000000000000} \label{Obj_alpha} \\
	 \Bigg[\begin{smallmatrix}
	c_{k1}^{i} \text{Tr} \{ \textbf{Y}_k^{i} W^{i}\} + a_k^{i}        & \sqrt{c_{k2}^{i}} \text{Tr} \{ \textbf{Y}_k^{i} W^{i}\} + b_k^{i} \\
	\sqrt{c_{k2}^{i}} \text{Tr} \{ \textbf{Y}_k^{i} W^{i}\} + b_k^{i} & -1                                                \\
	\end{smallmatrix} \Bigg] \preceq 0 \label{alpha}
\end{align}  
where $a_k^i : = - \alpha_k^{i} + c_{k0}^{i} + c_{k1}^{i} P_{D_k}^{i}$ and $b_k^{i} : = \sqrt{c_{k2}^{i}} P^{i}_{D_k}$. In addition, the apparent branch flow constraint \eqref{SlmCon} is rewritten:
\begin{align}
	\Bigg[ \begin{smallmatrix}
	- (\overline{S}_{lm}^{i})^2                & \text{Tr} \{ \textbf{Y}_{lm}^{i} W^{i}\} & \text{Tr} \{ \bar{\textbf{Y}}_{lm}^{i} W^{i}\} \\
	\text{Tr} \{ \textbf{Y}_{lm}^{i} W^{i}\}       & -1                               & 0                                      \\
	\text{Tr} \{ \bar{\textbf{Y}}_{lm}^{i} W^{i}\} & 0                                & -1
	\end{smallmatrix}\Bigg] \preceq 0
	\label{SlmConSDP}
\end{align}

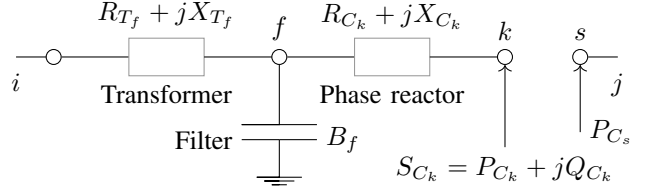
\begin{figure}
\center
\begin{tikzpicture}[
	font = \normalsize,
    dot/.style = {
      draw,
      fill = white,
      circle,
      inner sep = 2pt,
      minimum size = 0pt
    }
  ]
  \path (0,0) --++ (0,-0.5) coordinate (n3bl);
  \path (8.0,0) --++ (0,-0.6) coordinate[label = {below: $j$}] (hhhh);
   \path (-0.00,0) --++ (0,-0.6) coordinate[label = {below: $i$}] (hhhh);
  \path (7.5,0) --++ (0,-0.5) coordinate (n4bl);
  \draw (n3bl) --++ (0.5,0)coordinate[dot](n1bl);
  \draw (n4bl) --++ (0,0)coordinate[dot,label = {above: $s$}](s);
  \draw (s) --++ (0.5,0);
  \path (n1bl) --++ (1.5,0.25) coordinate[label = {above: $R_{T_f} + j X_{T_f}$}] (ymax);
  \path (n1bl) --++ (1.5,-0.25) coordinate[label = {below: Transformer}] (ymax);
  \draw (n1bl) --++ (1,0) coordinate(TFl);
  \draw [gray!80] (TFl) --++ (0,0.25) --++ (1,0) --++ (0,-0.5) --++ (-1,0) --++ (0,0.25);
  \path (TFl) --++ (1,0) coordinate (TFr);
  \draw (TFr) --++ (1,0) coordinate[dot,label = {above: $f$}](f);
  \draw (f) --++ (0,-0.9) --++ (-0.5,0) --++ (1,0);
  \path (f) --++ (0,-1.1) coordinate (fb);
  \draw (fb) --++ (-0.5,0) coordinate[label = {left: Filter}](ymax);
  \draw (fb) --++ (0,-0.5) --++ (-0.3,0) --++ (0.6,0);
  \path (fb) --++ (0,-0.55) coordinate (fbb);
  \draw (fbb) --++ (-0.2,0) --++ (0.4,0);
  \path (fb) --++ (0,-0.6) coordinate (fbbb);
  \draw (fbbb) --++ (-0.1,0) --++ (0.2,0);
  \draw (fb) --++ (0.5,0) coordinate[label = {right: $B_f$}](ymax);
  \draw (f) --++ (1,0) coordinate(PRl);
  \draw [gray!80] (PRl) --++ (0,0.25) --++ (1,0) --++ (0,-0.5) --++ (-1,0) --++ (0,0.25);
  \path (PRl) --++ (1,0) coordinate (PRr);
  \draw (PRr) --++ (1,0) coordinate[dot,label = {above: $k$}](k);
  \path (PRl) --++ (0.5,0.25) coordinate[label = {above: $R_{C_k} + j X_{C_k}$}] (ymax);
  \path (PRl) --++ (0.5,-0.25) coordinate[label = {below: Phase reactor}] (ymax);
  \draw [<-,black] (k) --++ (0,-1.2) coordinate[label = {below: $S_{C_k} = P_{C_k} + j Q_{C_k}$}] (ymax);
  \draw [<-,black] (s) --++ (0,-1) coordinate[label = {right: $P_{C_s}$}] (ymax);
\end{tikzpicture}
\vspace{-0.35cm}
\caption{Model of HVDC VSC connecting AC grid $i$ to HVDC grid $j$ \cite{beerten2012generalized}.}
\label{VSC_Schematic}
\vspace{-0.35cm}
\end{figure}
Fig.~\ref{VSC_Schematic} shows the model of the HVDC converter with filter bus $f$, AC bus $k$ and DC bus $s$ connecting AC grid $i$ to HVDC grid $j$. We model the HVDC converters as Voltage Source Converter (VSC) and make the following assumptions based on the work in \cite{bahrami2017semidefinite}: Each VSC  can control the active power $P_{C_k}$, and either the reactive power $Q_{C_k}$ or the AC terminal voltage. A transformer with resistance $R_{T_f}$ and reactance $X_{T_f}$ connects the AC grid to the filter bus $f$. The resistance $R_{C_k}$ of the phase reactor in the VSC is substantially smaller than its reactance $X_{C_k}$. The set of converters is denoted with the term~$\mathcal{C}$. The converter is able to modulate the voltage from the AC $i$ to the DC side $j$ by a certain modulation factor $m$:
\begin{align}
\text{Tr}\{\textbf{M}_k W^{i}\} & \leq m^2 \text{Tr} \{ \textbf{M}_s W^{j} \} \label{Mod_HVDC}  
\end{align}
The following active power balance has to hold between the AC bus $k$ and DC bus $s$:
\begin{align}
P_{C_k} + P_{C_s} + P^{\text{conv}}_{\text{loss},k} = 0 \label{PowerBalance}
\end{align}
The term $P^{\text{conv}}_{\text{loss},k}$ denotes the converter active power losses. To determine the exact converter losses a detailed assessment of the power electronic switching behavior is necessary, which substantially differs for each converter technology \cite{jones2013calculation}.
In this work, we model converter losses as a sum of a constant $a_k$ and a term that depends quadratically with factor $c_k$ on the converter current magnitude $|I_k|$: 
\begin{align}
P^{\text{conv}}_{\text{loss},k} = a_k +  c_k |I_k|^2  \label{ConvLoss}
\end{align} 
It is also possible to include an additional term which depends linearly on the converter current. As shown in \cite{bahrami2017semidefinite}, however, this requires to the introduction of a matrix variable containing the converter current and its squared value; and a relaxation of the rank constraint on this variable to achieve a convex formulation. A penalization term is then required in the objective function to enforce the rank-1 property for this matrix variable. As this complicates the formulation, we choose here to use a loss model that uses a constant and quadratic term. With Ohm's Law, the current flow magnitude $|I_k|$ from filter bus $f$ to AC bus $k$ of the converter at AC side $i$ is:
\begin{align}
|I_k|^2 = (R_{C_k}^2 + X_{C_k}^2)^{-1} \text{Tr} \{ \textbf{M}_{kf} W^{i}\} \label{CurrFlow}
\end{align}
The converter current $|I_k|^2$ from \eqref{CurrFlow} can be inserted in the converter power balance \eqref{PowerBalance} using the converter power losses \eqref{ConvLoss} with $z_{C_k} := c_k (R_{C_k}^2 + X_{C_k}^2)^{-1}$:
\begin{align}
\text{Tr} \{ \textbf{Y}_k W^{i}\} + \text{Tr} \{ \textbf{Y}_s W^{j}\} & + a_k +    \nonumber \\
z_{C_k} \text{Tr} \{ \textbf{M}_{kf} W^{i}\} & + P_{D_k}^{i} + P_{D_s}^{i} = 0 \label{Conv_Pbal} 
\end{align} 
\begin{figure}
\center
\begin{tikzpicture}[
	font = \normalsize,
    thick,
    >=stealth',
    dot/.style = {
      draw,
      fill = gray!40,
      circle,
      inner sep = 2pt,
      minimum size = 0pt
    }
  ]
  \draw[white,fill=blue!40] (0,0) ellipse (1.5 and 1.5);
  \draw[white,fill=white] (-2,0.6) -- (+2.5,0.6) -- (4,2) -- (-4,2);
  \draw[white,fill=white] (-2,-0.8) -- (+2.5,-0.8) -- (4,-1.55) -- (-4,-1.55);
  \draw[blue,dashed] (0,0) ellipse (1.5 and 1.5);
  \draw[->] (-2,0) -- (2,0) coordinate[label = {below:$P_{C_k}$}] (xmax);
  \draw[->] (0,-1.65) -- (0,1.8) coordinate[label = {right:$Q_{C_k}$}] (ymax);
  \draw[blue,dashed] (-2,0.6) -- (2,0.6) coordinate[label = {right:$m_c S_{C_k}^{\text{nom}}$}] (ymax);
  \draw[blue,dashed] (-2,-0.8) -- (2,-0.8) coordinate[label = {right:$- m_b S_{C_k}^{\text{nom}}$}] (ymax);
	  \draw[blue] plot [smooth] coordinates {(1,1.1180) (1.2,1.4) (1.5,1.2) (1.7,1.5)};
	  \path[blue] (0,0) -- (1.8,1.4)coordinate[label = {above:$|V_k| \overline{I}_k$}] (ymax);
	  \draw[black] plot [smooth] coordinates {(-1.2,0.2) (-1.8,0.4) (-2.0,1.0)};
	  \path[black] (0,0) -- (-2.5,0.8) coordinate[label = {above: $\begin{matrix} \text{Feasible} \\ \text{operating region} \end{matrix}$}] (ymax);
\end{tikzpicture}
\vspace{-0.2cm}
\caption{Active and reactive power capability curve of HVDC converter \cite{imhof2015voltage}.}
\vspace{-0.35cm}
\label{PQ_Conv}
\end{figure}
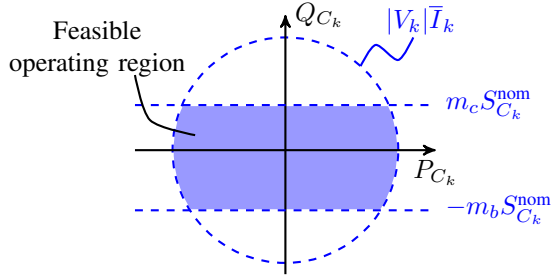

The converter has a feasible operating region to inject and absorb both reactive and active power as depicted in Fig.~\ref{PQ_Conv}. The maximum reactive power which can be absorbed or injected by the converter is lower- and upper-bounded with positive constants $m_b$ and $m_c$ as follows \cite{imhof2015voltage}:
\begin{align}
-m_b S_{C_k}^{\text{nom}}  & \leq \text{Tr} \{\bar{\textbf{Y}}_k  W\}  \leq m_c S_{C_k}^{\text{nom}}  \label{Conv_S1}
\end{align} 
The nominal apparent power rating of the converter is given by $S_{C_k}^{\text{nom}}$. The   maximum transferable apparent power $S_{C_k}$ is upper-bounded by the converter current limit $\overline{I}_k$:
\begin{align}
|S_{C_k}|^2 = (P_{C_k})^2 + (Q_{C_k})^2 \leq (|V_k| \overline{I}_k)^2 \label{S_conv_up}
\end{align}
The constraint on the apparent branch flow through the converter \eqref{S_conv_up} is rewritten using Schur's complement:
\begin{align}
\Bigg[ \begin{smallmatrix}
\overline{I}_k^2 \text{Tr} \{ \textbf{M}_k W^{i} \} & \text{Tr} \{ \textbf{Y}_k^{i} W^{i}\} + P_{D_k}^{i} & \text{Tr} \{ \bar{\textbf{Y}}_k^{i} W^{i}\} + Q^{i}_{D_k} \\
\text{Tr} \{ \textbf{Y}_k^{i} W^{i}\} + P^{i}_{D_k} &  1 & 0 \\
\text{Tr} \{ \bar{\textbf{Y}}_k^{i} W^{i}\} + Q^{i}_{D_k} & 0 & 1 \\
\end{smallmatrix} \Bigg] \succeq 0 \label{S_convM}
\end{align}
The non-convex AC-OPF minimizes the objective \eqref{Obj_alpha} subject to AC and HVDC grid constraints \eqref{PBal} -- \eqref{PlmCon}, \eqref{WTT}, \eqref{alpha}, \eqref{SlmConSDP}, and HVDC converter constraints \eqref{Mod_HVDC}, \eqref{Conv_Pbal} -- \eqref{Conv_S1}, \eqref{S_convM}. The non-convex rank constraint \eqref{WTT} can be expressed by:
\begin{align}
	W^{i} \succeq 0 \label{SDP}         \\
	\text{rank}(W^{i}) = 1 \label{Rank}
\end{align}
The convex relaxation is introduced by dropping the rank-1 constraint \eqref{Rank}, relaxing the non-linear, non-convex AC-OPF to a convex semidefinite program (SDP). The work in \cite{lavaei2012zero} proves for AC grids that if the rank of $W$ obtained from the SDP relaxation is 1 or 2, then $W$ is the global optimum of the non-linear, non-convex AC-OPF and the optimal voltage vector can be computed following the procedure described in \cite{molzahn2013implementation}. Whether the rank is 1 or 2 when the relaxation is exact, depends on if the slack bus angle is fixed as an additional constraint in the AC-OPF. In \cite{bahrami2017semidefinite}, there are two necessary conditions to obtain zero relaxation gap for interconnected AC and HVDC grids: First, as in \cite{lavaei2012zero}, a small resistance of $10^{-4} \, \text{p.u.}$ has to be included for each transformer. Second, a large resistance of $10^4 \, \text{p.u.}$ has to be added between the AC bus $k$ and DC bus $s$ of each converter. This is to ensure the resistive connectivity of the power system graph. In our work, we eliminate the need for the second condition. We formulate the problem not as a function of one matrix $W$ for the whole grid, as in \cite{bahrami2017semidefinite}, but one matrix $W^{i}$ for each AC and HVDC grid. This allows us to eliminate the need for the large resistance and still obtain zero relaxation gap. This leads to two desirable effects. First, the numerical stability is increased as the high value of $10^4 \, \text{p.u.}$ used to be causing numerical scaling problems to the SDP solver in our experiments. Second, the computational run time is reduced, as we consider a reduced amount of matrix entries.

In order to further increase scalability, a chordal decomposition of the semidefinite constraints is applied. Following \cite{jabr2012exploiting}, in order to obtain a chordal graph, a chordal extension of each AC and HVDC grid graph is computed with a Cholesky factorization. Then, we compute the maximum cliques decomposition of the obtained chordal graph. We replace the semidefinite constraint \eqref{SDP} with:
\begin{align}
(W^{i})_{clq,clq} \succeq 0 \label{chordal}
\end{align} 
The positive semidefinite matrix completion theorem ensures that if \eqref{chordal} holds for each maximum clique $clq$, the resulting matrix $W^{i}$ can be completed such that it is positive semidefinite. This allows to substantially reduce the number of considered matrix entries and the computational burden \cite{jabr2012exploiting}. The chordal decomposition requires additional equality constraints between matrix entries of $W$ appearing in several cliques to ensure consistency. There is a computational tradeoff between the complexity of the decomposed semidefinite constraint \eqref{chordal} and  the number of those equality constraints. Using  heuristic clique merging, an optimal computational trade-off can be achieved \cite{molzahn2013implementation, fukuda2001exploiting}.
\subsection{Inclusion of Chance Constraints}
Renewable energy sources and stochastic loads introduce uncertainty in power system operation. To account for uncertainty in bus power injections, we extend the presented OPF formulation with a joint chance constraint. A total number of $n_W$ wind farms are introduced in the system of interconnected AC and HVDC grids at buses $k \in \mathcal{W}$ and modeled as
\begin{equation}
	P_{W_k} = P_{W_k}^f + \zeta_k \quad,
\end{equation}
where $P_{W}$ are the actual wind infeeds, $P_{W_k}^f$ are the forecasted values and $\zeta$ are the uncertain forecast errors. 
The chance constrained AC-OPF for interconnected AC and HVDC grids uses the semidefinite relaxation of the AC-OPF and includes a joint chance constraint for all buses $k \in \mathcal{N}^{i}$, lines $(l,m) \in \mathcal{L}^{i}$ and converters $(s,k,f,i,j) \in \mathcal{C}$:
\begin{align}
	& \min_{\alpha^{i}, W_0^{i}, W^{i}(\zeta)} \, \, \sum_{i=0}^{n_{\text{grid}}} \sum_{k \in \mathcal{G}^{i}} \alpha_k^{i}  \label{ObjCh} \\
	\text{s.t.\, }     &  \text{\eqref{alpha}, \eqref{PBal} -- \eqref{PlmCon}, \eqref{SlmConSDP}, \eqref{Mod_HVDC}, \eqref{Conv_Pbal}, \eqref{Conv_S1}, \eqref{S_convM}, \eqref{chordal}} \nonumber   \\
	& \text{  for }   W^{i} = W^{i}_0 \label{AllCon}     \\
	&\mathbb{P} \Big\{ \text{\eqref{PBal} -- \eqref{PlmCon}, \eqref{SlmConSDP}, \eqref{Mod_HVDC}, \eqref{Conv_Pbal}, \eqref{Conv_S1}, \eqref{S_convM}, \eqref{chordal}} \nonumber  \Big\} \geq 1-\epsilon  \\
	&   \text{   for } W^{i} = W^{i}(\zeta) \label{SDPCh}
\end{align}
For a given maximum allowable violation probability $\epsilon \in (0,1)$, the joint chance constraint \eqref{SDPCh} ensures that compliance with the system constraints  is achieved with probability higher than the confidence interval $1-\epsilon$. The formulation with a joint chance constraint is desirable from the operator's point of view, as it ensures for a given probability that the entire AC and HVDC system state is secured against the uncertainty. The system constraints can be classified in two types. The constraints corresponding to equations \eqref{PBal}--\eqref{PlmCon}, \eqref{Mod_HVDC}, \eqref{Conv_Pbal} -- \eqref{Conv_S1} are linear scalar constraints and those corresponding to equations \eqref{SlmConSDP}, \eqref{S_convM}, \eqref{chordal} are semi-definite constraints. The matrix $W_0^{i}$ is the forecasted system state and the matrix $W^{i}(\zeta)$ denotes the system state as a function of the forecast errors which are the continuous uncertain variables $\zeta$. Hence, the chance constrained AC-OPF problem \eqref{ObjCh} -- \eqref{SDPCh} is an infinite-dimensional problem optimizing over $W^{i}(\zeta)$  \cite{vrakopoulou2013probabilistic}. This renders the problem intractable, which makes it necessary to identify a suitable approximation for $W^{i}(\zeta)$ \cite{ben2008selected}.  In the following, an approximation of an explicit dependence of $W^{i}(\zeta)$ on the forecast errors is presented. 
\section{Tractable Optimal Power Flow Formulation} \label{III}
 Using a scenario-based method, we define the uncertainty set associated with forecast errors. As the optimization problem is infinite-dimensional, we use a piecewise affine approximation to model the system state as a function of forecast errors. This allows us to include corrective control policies of active and reactive power set-points of HVDC converters, and of generator active power and voltage set-points. To achieve tractability of the resulting chance constraint, theoretical results from robust optimization are leveraged. By using a penalty term on power losses, we introduce a heuristic method to identify suitable penalty terms to obtain rank-1 solution matrices. We show how the proposed AC-OPF formulation can be decomposed using Benders decomposition.
\subsection{Scenario-Based Uncertainty Set} \label{III.1}
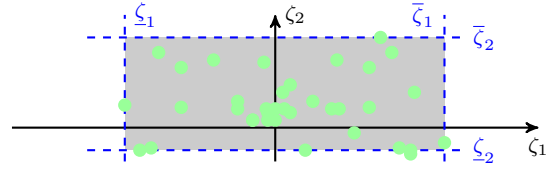
\begin{figure}
\center
\begin{tikzpicture}[
	font = \footnotesize,
    thick,
    >=stealth',
    dot/.style = {
      draw,
      fill = green!40,
      circle,
      inner sep = 1.5pt,
      minimum size = 0pt
    }
  ]
  \coordinate (O) at (0,0);
  \fill[gray!40]  (-2,-0.3) -- (-2,1.2) -- (2.25,1.2) -- (2.25,-0.3) --  (-2,-0.3);
  \draw[->] (-3.5,0) -- (3.5,0) coordinate[label = {below: $\zeta_1$}] (xmax);
  \draw[->] (0,-0.45) -- (0,1.5) coordinate[label = {right: $\zeta_2$}] (ymax);
      \draw[blue,dashed] (-2,-0.45) -- (-2,1.5) coordinate[label = {right: $\underline{\zeta}_1$}] (ymax);
 	\draw[blue,dashed] (-2.5,1.2) -- (2.5,1.2) coordinate[label = {right: $\overline{\zeta}_2$}] (ymax);
 	\draw[blue,dashed] (2.25,-0.45) -- (2.25,1.5) coordinate[label = {left: $\overline{\zeta}_1$}] (ymax);
 	\draw[blue,dashed] (-2.5,-0.3) -- (2.5,-0.3)coordinate[label = {right: $\underline{\zeta}_2$}] (ymax);
  \node[dot,green!40] at (0,0.1){};
  \node[dot,green!40] at (0,0.25){};
  \node[dot,green!40] at (0.12,0.25){};
  \node[dot,green!40] at (0.82,0.9){};
  \node[dot,green!40] at (-0.1,0.1){};
  \node[dot,green!40] at (-0.5,0.25){};
    \node[dot,green!40] at (-0.1,0.2){};
  \node[dot,green!40] at (-0.5,0.35){};
    \node[dot,green!40] at (-0.3,0.1){};
  \node[dot,green!40] at (0.75,0.25){};
      \node[dot,green!40] at (0.2,0.2){};
  \node[dot,green!40] at (0.5,0.35){};
    \node[dot,green!40] at (-0.3,0.1){};
  \node[dot,green!40] at (0.85,0.25){};
  \node[dot,green!40] at (-0.12,0.25){};
  \node[dot,green!40] at (-0.82,0.9){};
  \node[dot,green!40] at (-2,0.3){};
  \node[dot,green!40] at (1.8,-0.35){};
  \node[dot,green!40] at (0.4,-0.3){};
  \node[dot,green!40] at (1.4,1.2){};
  \node[dot,green!40] at (2.25,-0.2){};
    \node[dot,green!40] at (-1.8,-0.3){};
    \node[dot,green!40] at (-1.55,1.0){};
        \node[dot,green!40] at (-1.25,0.8){};
     \node[dot,green!40] at (-1.25,0.27){};
         \node[dot,green!40] at (-1.65,-0.27){};
    \node[dot,green!40] at (1.8,-0.3){};
    \node[dot,green!40] at (1.55,1.0){};
        \node[dot,green!40] at (1.25,0.8){};
     \node[dot,green!40] at (1.25,0.27){};
         \node[dot,green!40] at (1.65,-0.27){};
              \node[dot,green!40] at (1.85,0.47){};
         \node[dot,green!40] at (1.05,-0.07){};
                       \node[dot,green!40] at (0.1,0.47){};
                                 \node[dot,green!40] at (0.2,0.57){};
         \node[dot,green!40] at (-0.15,0.87){};
\end{tikzpicture}
\vspace{-0.25cm}
\caption{Illustration how the bounds on the forecast errors for two wind farms are retrieved using \cite{margellos2014road}. The green circles represent $N_{\text{s}}$ scenarios.}
\label{Scenario}
\end{figure}
To determine the bounds of the uncertainty set for a defined $\epsilon$, we use a scenario-based method from \cite{margellos2014road}, which does not make any assumption on the underlying distribution of the forecast errors. To this end, we compute the minimum volume rectangular set which with probability $1-\beta$ contains $1-\epsilon$ of the probability mass. The term $\beta$ is a confidence parameter which is usually initially selected to be very low. According to \cite{margellos2014road},  which builds upon \cite{campi2008exact}, it is necessary to draw at least the following number of scenarios $N_{\text{s}}$ to specify the uncertainty set:
	\begin{align}
		N_{\text{s}} \geq \dfrac{1}{\epsilon} \dfrac{e}{e-1} (\text{ln} \dfrac{1}{\beta} + 2 n_{\delta} - 1) \label{NumberOfSamples}
	\end{align}
The term $e$ is Euler's number and the term $n_{\delta}$ is the number of uncertain variables, which in our case is the number of wind farms $n_W$. The minimum and maximum bounds on the forecast errors $\zeta$ are retrieved by a simple sorting operation among the $N_{\text{s}}$ scenarios as illustrated in Fig.~\ref{Scenario}. The resulting rectangular uncertainty set has a number of $n_v$ vertices $v \in \mathcal{V}$ which are its corner points. For each vertex  $v \in \mathcal{V}$, the vector $\overline{\zeta}_v \in \mathbb{R}^{n_W}$ contains the corresponding forecast error magnitudes of the wind farms. Alternatively, the user could specify a polyhedral uncertainty set, or a number of scenarios to be included.
\subsection{Piecewise Affine Approximation} \label{III.2}
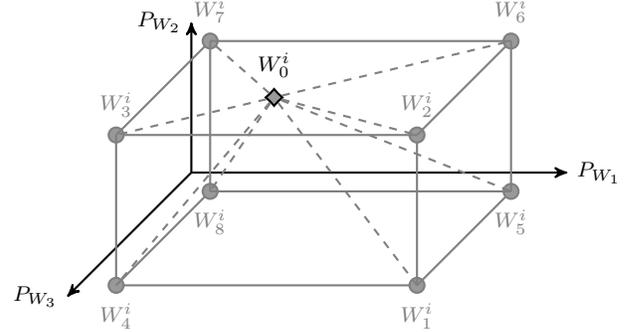
\begin{figure}
	\center
	\begin{tikzpicture}[
		font = \footnotesize,
		thick,
		>=stealth',
		dot/.style = {
			draw,
			fill = gray!80,
			circle,
			inner sep = 2pt,
			minimum size = 0pt
		},
		dott/.style = {
			draw,
			fill = gray!80,
			star,
			inner sep = 1.5pt,
			minimum size = 0pt,
			star points=4
		}
		]
	    \draw[->] (0,0) -- (5,0) coordinate[label = {right:$P_{W_1}$}] (xmax);
		\draw[->] (0,0) -- (0,2.0) coordinate[label = {left:$P_{W_2}$}] (ymax);
		\draw[->] (0,0) -- (-1.65,-1.65) coordinate[label = {left:$P_{W_3}$}] (zmax);
		\draw[gray] (-1,-1.5) -- (3,-1.5) node[dot,label = {below:$W_1^{i}$}]{};
		\draw[gray] (3,-1.5) -- (3,0.5) node[dot,label = {above:$W_2^{i}$}]{};
		\draw[gray] (3,0.5) -- (-1,0.5) node[dot,label = {above:$W_3^{i}$}]{};
		\draw[gray] (-1,0.5)-- (-1,-1.5) node[dot,label = {below:$W_4^{i}$}]{};
		\draw[gray] (-1,-1.5) -- (0.25,-0.25) node[dot,label = {below:$W_8^{i}$}]{};
		\draw[gray] (3,-1.5) -- (4.25,-0.25) node[dot,label = {below:$W_5^{i}$}]{};
		\draw[gray] (3,0.5) -- (4.25,1.75) node[dot,label = {above:$W_6^{i}$}]{};
		\draw[gray] (-1,0.5)-- (0.25,1.75) node[dot,label = {above:$W_7^{i}$}]{};
		\draw[gray] (0.25,1.75) -- (0.25,-0.25);
		\draw[gray] (0.25,-0.25) -- (4.25,-0.25);
		\draw[gray] (4.25,-0.25) -- (4.25,1.75);
		\draw[gray] (4.25,1.75)-- (0.25,1.75);
		\path(4.25,1.75)-- (1.1,1.0) coordinate[dott,label = {above:$W_0^{i}$}](W0);
		\draw[gray,dashed] (W0) -- (3,-1.5) ;
		\draw[gray,dashed] (W0) -- (3,0.5) ;
		\draw[gray,dashed] (W0) -- (-1,0.5) ;
		\draw[gray,dashed] (W0)-- (-1,-1.5) ;
		\draw[gray,dashed] (W0) -- (0.25,-0.25) ;
		\draw[gray,dashed] (W0) -- (4.25,-0.25)  ;
		\draw[gray,dashed] (W0) -- (4.25,1.75)  ;
		\draw[gray,dashed] (W0) --  (0.25,1.75)  ;
	\end{tikzpicture}
	\vspace{-12pt}
	\caption{Uncertainty set derived from the scenario-based method for three wind farms. It is sufficient to enforce the joint chance constraint at the vertices, i.e. at the obtained maximum bounds on forecast errors. For each grid $i$, the piecewise affine approximation interpolates the system state between the forecasted system states $W^{i}_0$ and the vertices $W^{i}_{1-8}$ denoted with circles.}
	\label{RectSet}
\end{figure}
For the previously obtained rectangular uncertainty set, we use the piecewise affine approximation from \cite{venzke2017} to model the system change as a function of the forecast errors. To this end, we introduce a matrix $W_v^{i}$ for each vertex $v \in \mathcal{V}$ and power grid $i$. The system state of each power grid $i$ as a function of the forecast errors is approximated as a piecewise affine interpolation between the forecasted system state $W_0^{i}$ and the vertices of the uncertainty set $W_v^{i}$: 
\begin{align}
	W^{i}(\zeta) & := W_0^{i} + \Psi_{v=1}^{n_\text{v}} (\zeta) (W^{i}_v-W^{i}_0) \nonumber 
\end{align}
The function $\Psi_{v=1}^{n_\text{v}} (\zeta) $ denotes a piecewise affine interpolation operator of the wind forecast error $\zeta$ between all vertices $\overline{\zeta}_v$. It returns a weight for the direction of each vertex, corresponding to the distance. For the case of three uncertain wind infeeds, this concept is illustrated in Fig.~\ref{RectSet}. 
\subsection{Corrective Control Policies} \label{III.3}
The piecewise affine approximation allows to include corresponding piecewise corrective control policies. We assume that the system operator can respond to forecast errors with corrective control of HVDC converters and generator active power and voltage set-points, i.e. the system operator sends updated set-points based on the realization of forecast errors. 

During steady-state power system operation, generation has to match demand and system losses. If an imbalance occurs due to e.g. an occurring forecast error, designated generators in the power grid will respond by adjusting their active power output as part of automatic generation control (AGC). The vector $d_G^{i} \in \mathbb{R}^{n_b^{i}}$ defines the generator participation factors for each grid $i$. The term $n_b^{i}$ denotes the number of buses of grid $i$. The vector $d^{i}_{w} \in \mathbb{R}^{n_b^{i}}$ has a $\{-1\}$ entry corresponding to the bus where the $w$-th wind farm is located. The other entries of this vector are $\{0\}$. The sum of the generator participation factors should compensate the deviation in wind generation, i.\,e.\,$ \sum_{i}^{n_{\text{grid}}} \sum_{k \in G} d^{i}_{G_k} = 1$. The line losses of the AC power grid vary non-linearly with changes in wind infeeds and corresponding adjustments of generator output. To allow for a compensation of the change in system losses, we introduce a slack variable $\gamma_v$ for each vertex. In order to link the generation dispatch of the forecasted system state $W_0^{i}$ with system states in which forecast errors occur, the following constraints are introduced for each vertex $v \in \mathcal{V}\backslash\{0\}$, bus $k \in \mathcal{N}^i$ and power grid $i$:
\begin{align}
	\text{Tr} \{ \textbf{Y}^{i}_k (W_v^{i}  -  W_0^{i}) \}   =\sum_{w=1}^{n_W} & \overline{\zeta}_{v_w} (d_{G_k}^{i} + \gamma_v d_{G_k}^{i} + d^{i}_{w_k}) 
\end{align}
As a result, it is ensured that each generator compensates the non-linear change in system losses proportional to its participation factor. We allow for a corrective control of voltage set-points at generator terminals in case of forecast errors and recover the set-point at bus  $k \in \mathcal{N}^i$ and grid $i$:
\begin{align}
    |\mathbf{V}^{i}_k|^2 := \text{Tr} \{ \textbf{M}_k W_0^{i} \} + \Psi_{v=1}^{n_\text{v}} (\zeta)  \text{Tr} \{ \textbf{M}_k (W^{i}_v-W^{i}_0) \} \nonumber
\end{align}

Grid codes specify reactive power capabilities of wind farms often in terms of power factor $\cos \phi = \sqrt{\tfrac{P^2}{P^2+Q^2}}$. We allow for a power factor set-point being sent to each wind farm. Note that our AC-OPF framework captures the variation of the wind farm reactive power injection as a function of wind farm active power. To this end, we modify constraints \eqref{QBal} to include the reactive power capabilities of wind farms. We introduce the reactive power set-point $\tau_k$ for each wind farm $k \in \mathcal{W}$:
\begin{align}
-\sqrt{\tfrac{1-\cos^2\phi}{\cos^2 \phi}} \leq \tau_k \leq \sqrt{\tfrac{1-\cos^2\phi}{\cos^2 \phi}} \label{QCor}
\end{align}

The HVDC converter can be operated in $PV$ or $PQ$ control mode. Here, we consider the latter, that is the HVDC converter is able to independently control its active and reactive power set-point. Note that our framework can capture the $PV$ control mode as well. The resulting controllability can be utilized to react to uncertain power injections by rerouting active power flows, injecting, or absorbing additional reactive power. In this work, a piecewise affine corrective control policy is introduced for the HVDC converter. The optimization determines an optimal set-point for the active and reactive power of the converter for each vertex and for the operating point. The set-points for a realization of the forecast errors $\zeta$ are computed as a piecewise affine interpolation for converter $k \in \mathcal{C}$:
\begin{align}
P_{C_k} (\zeta)  := \text{Tr} \{\textbf{Y}_k^{i}  W_0^{i} \} & - P^{i}_{D_k} +  \Psi_{v=1}^{n_\text{v}} (\zeta)  \text{Tr} \{ \textbf{Y}_k^{i} (W^{i}_v-W^{i}_0) \}  \nonumber \\
Q_{C_k} (\zeta)  := \text{Tr} \{\bar{\textbf{Y}}_k^{i}  W_0^{i} \} & - Q^{i}_{D_k} + \Psi_{v=1}^{n_\text{v}} (\zeta)  \text{Tr} \{ \bar{\textbf{Y}}_k^{i} (W^{i}_v-W^{i}_0) \}  \nonumber 
\end{align}
\subsection{Robust Optimization} \label{III.4}
To obtain a tractable formulation of the chance constraints including the control policies, the following result from robust optimization  is used: If the constraint functions are linear, monotone or convex with respect to the uncertain variables, then the system variables will take the maximum values at the vertices of the uncertainty set \cite{margellos2014road}. Using the piecewise affine approximation of Section~\ref{III.2}, the system constraints corresponding to equations \eqref{PBal}--\eqref{PlmCon}, \eqref{Mod_HVDC}, \eqref{Conv_Pbal} -- \eqref{Conv_S1} are linear and those corresponding to equations \eqref{SlmConSDP}, \eqref{S_convM}, \eqref{chordal} are semidefinite, i.e. convex. Hence, it suffices to enforce the joint chance constraint at the vertices $v \in \mathcal{V}$ of the rectangular uncertainty set or the corner points of a polyhedral uncertainty set. We provide a tractable formulation of \eqref{SDPCh} for each vertex $v \in \mathcal{V}$, bus $k \in \mathcal{N}^i$, line $(l,m) \in \mathcal{L}^i$ and grid $i$; the converter constraints are formulated for each converter $(s,k,f,i,j) \in \mathcal{C}$:
\begin{align}
	& 	\underline{P}^{i}_k  \leq \text{Tr}\{ \textbf{Y}^{i}_k W_v^{i} \} + P^{i}_{D_k} - P_{W_k}^f - \overline{\zeta}_{v_k} \leq \overline{P}^{i}_k \label{PnodalRect} \\
	&	\underline{Q}^{i}_k    \leq \text{Tr}\{ \bar{\textbf{Y}}^{i}_{k} W_v^{i}  \}  + Q^{i}_{D_k} - \tau_k (P^f_{W_k} + \overline{\zeta}_{v_k}) \leq \overline{Q}^{i}_k    \\
	&	(\underline{V}^{i}_k)^2  \leq  \text{Tr}\{ M_{k} W_v^{i} \}   \leq (\overline{V}^{i}_k)^2     \label{VnodalRect}                        \\
	&	- \overline{P}^{i}_{lm} \leq \text{Tr}\{ \textbf{Y}^{i}_{lm} W_v^{i} \}  \leq \overline{P}^{i}_{lm}\\
	&	\Bigg[ \begin{smallmatrix}
	- (\overline{S}_{lm}^{i})^2                   & \text{Tr} \{ \textbf{Y}^{i}_{lm} W_v^{i}  \} & \text{Tr} \{ \bar{\textbf{Y}}^{i}_{lm} W_v^{i}   \} \\
	\text{Tr} \{ \textbf{Y}^{i}_{lm} W_v^{i}   \}      & -1                                  & 0                                          \\
	\text{Tr} \{ \bar{\textbf{Y}}^{i}_{lm} W_v^{i}  \} & 0                                   & -1
	\end{smallmatrix} \Bigg] \preceq 0 \label{Slm_Rect}\\
		& W^{i}_{v,(clq,clq)} \succeq 0 \label{SDPRect} \\
& \text{Tr} \{ \textbf{Y}^{i}_k W_v^{i}\} + \text{Tr} \{ \textbf{Y}^{i}_s W_v^{i}\} + a_k +  \nonumber \\ 
& \quad  \quad \quad \quad \quad    \quad  z_{C_k} \text{Tr}  \{ \textbf{M}_{kf}  W^{i}\} + P_{D_k}^{i} + P_{D_s}^{i} = 0 
 \label{R_Conv_Vol} \\
 & \text{Tr} \{ \textbf{M}_k W_v^{i}\} \leq m^2 \text{Tr} \{ \textbf{M}_s W_v^{j}\} \\
 & -m_b S_{C_k}^{\text{nom}}  \leq  \text{Tr} \{ \bar{\textbf{Y}}_k  W_v^{i}\}  \leq m_c S_{C_k}^{\text{nom}}  \label{R_Conv_Low_lim} \\
 & \left[ \begin{smallmatrix}
\overline{I}_k^2 \text{Tr} \{\textbf{M}_k  W_v^{i}\}  & \text{Tr}\{ \textbf{Y}^{i}_{k} W_v\} + P^{i}_{D_k}   & \text{Tr}\{ \bar{\textbf{Y}}^{i}_{k} W_v\} + Q^{i}_{D_k}  \\
\text{Tr}\{ \textbf{Y}_{k}^{i} W^{i}_v\} + P^{i}_{D_k} &  1 & 0 \\
\text{Tr}\{ \bar{\textbf{Y}}^{i}_{k} W^{i}_v\} + Q^{i}_{D_k} & 0 & 1 \\
\end{smallmatrix} \right]  \succeq 0 \label{R_Conv_Up_lim_AP}
\end{align}
\begin{align}
			&	\text{Tr} \{ \textbf{Y}^{i}_k (W_v^{i}  -  W_0^{i}) \}   =\sum_{w}^{n_W}  \zeta_{v_w} (d_{G_k}^{i} + \gamma_v d_{G_k}^{i} + d^{i}_{w_k}) \label{LinkHVDC2}
	\end{align}
Equation \eqref{LinkHVDC2} links the forecasted system state with each of the vertices $v$. The chance constrained AC-OPF formulation minimizes \eqref{ObjCh} subject to \eqref{AllCon}, \eqref{QCor}, \eqref{PnodalRect} --  \eqref{LinkHVDC2}. 
\subsection{Systematic Procedure to Obtain Rank-1 Solution Matrices}\label{III.5}
In case we do not obtain rank-1 solution matrices for all vertices of the uncertainty set, we propose to add an active power loss penalty term to the objective function \eqref{ObjCh}, where the terms $\mu_v \geq 0$ are weighting factors:
\begin{align}
	\min_{\alpha^{i}, W_0^{i}, W_v^{i}, \tau_k, \gamma_v} \, \, \sum_{i=0}^{n_{\text{grid}}} \sum_{k \in \mathcal{G}^{i}} \alpha_k^{i}  + \sum_{v=1}^{n_v} \mu_v \gamma_v  \label{ObjPen}
\end{align}
We use an individual penalty term $\mu_v$ for each vertex and outage instead of a uniform penalty parameter $\mu$ as in \cite{venzke2017}. We found in \cite{venzke2018} that this allows us to introduce a robust systematic method to identify rank-1 solution matrices, as we will show in Section~\ref{IV}. For this purpose, we solve the chance constrained AC-OPF in an iterative manner. First, we set all penalty weights  $\mu_v$ to 0 and solve the OPF problem. If we obtain rank-1 $W$ solutions, we terminate. Otherwise, we increase the penalty weight $\mu_v$ by a defined step-size $\Delta \mu$ only for higher rank matrices and re-solve the OPF problem. We repeat this procedure until all $W$ matrices are rank-1 (in each grid $i$, there is a $W$ matrix for the operating point, and one additional $W$ matrix for each vertex, see Fig. \ref{RectSet}). With the penalized semidefinite AC-OPF formulation, near-global optimality guarantees can be derived, which can specify the maximum distance to the global optimum of a non-convex AC-OPF using the piecewise affine approximation \cite{madani2015convex}. The numerical results in Section~\ref{IV.2} show that while this penalty is necessary to obtain rank-1 solution matrices, in practice the deviation of the near-global optimal solution \eqref{ObjPen} from the global optimum is small. Alternatively, a penalty term on the generator reactive power injections based on \cite{madani2015convex} can be introduced for each vertex in the objective function:
\begin{align}
	\min_{\alpha^{i}, W_0^{i}, W_v^{i}, \tau_k, \gamma_v} \, \, \sum_{i=0}^{n_{\text{grid}}} \sum_{k \in \mathcal{G}^{i}} \alpha_k^{i}  +  \sum_{v=1}^{n_v} \mu_v \sum_{i=0}^{n_{\text{grid}}} \sum_{k \in \mathcal{G}^{i}} Q_{G_k}^{v,i} \label{ObjPenQ}
\end{align}
The term $  Q_{G_k}^{v,i}$ denotes the reactive power injection at bus $k$ of grid $i$ and vertex $v$. The objective \eqref{ObjPenQ} minimizes generation cost and penalizes the generator reactive power injections for each vertex. We show in our simulation studies in Section~\ref{IV} that the reactive power penalty term can also obtain rank-1 solution matrices but leads to substantially larger upper bounds for the distance to the global optimum than the active power loss penalty \eqref{ObjPen}, i.e. a higher generation cost.
\subsection{Benders Decomposition For Vertices of Uncertainty Set} \label{III.6} 
In the following, using Benders decomposition, we show how the proposed optimization problem can be decomposed in one master problem, and one subproblem for each vertex of the uncertainty set. This is desirable as the number of vertices in the proposed OPF formulation grows exponentially with the number of wind farms. For a detailed explanation of decomposition techniques the interested reader is referred to \cite{birge2011introduction}. We assume the power factor $\tau$ of the wind farms is fixed. Then, the forecasted system state $W_0$ couples the system states for each vertex $W_v$ only through the equality constraints \eqref{LinkHVDC2}. In fact, only the active generator power dispatch $P_{G,0}$ of the forecasted system state links the vertices, i.e. is the complicating variable. If $P_{G,0}$ is assumed fixed, then the optimization problem decomposes into one subproblem for each vertex. The master problem at iteration $J$ can be stated:
\begin{align}
      	&  \min_{\alpha^{i}, W_0^{i}, P_{G,0}^i, \Theta \geq \Theta_{\text{min}}} \, \, \, \, \sum_{i=0}^{n_{\text{grid}}} \sum_{k \in \mathcal{G}^{i}} \alpha_k^{i}  + \Theta \label{ObjMas} \\
	    \text{s.t.\, }   & \eqref{AllCon}, P_{{G_k},0}^i = \text{Tr} \{ \textbf{Y}^{i}_k W_0^{i} \} + P_{D_k}^i \quad \forall k \in \mathcal{G}^{i}    \\ 
    	& \Theta \geq \sum_{v=1}^{n_v}  \mu_v \gamma_v^{(j)} + \sum_{i=0}^{n_{\text{grid}}}  \sum_{k \in \mathcal{G}^{i}} \Lambda_{k}^{v,(j)} (P_{{G_k},0}^i - P_{{G_k},0}^{i,(j)}) \nonumber \\
    	& \quad \quad \quad \quad \forall j = 1, ..., J-1 \label{OPT_CUTS} \\
    	& 0 \geq  S_v+ \sum_{i=0}^{n_{\text{grid}}}  \sum_{k \in \mathcal{G}^{i}}\Omega_{k}^{v,(j)} (P_{{G_k},0}^i - P_{{G_k},0}^{i,(j)})   \nonumber \\
    	& \quad \quad \quad \quad \forall v \in \mathcal{V},\, j = 1, ..., J-1 \label{FEAS_CUTS} 
\end{align}
The original objective function is reconstructed using the optimality cuts \eqref{OPT_CUTS} for the auxiliary variable $\Theta$. In case subproblems are infeasible for the chosen generation dispatch $P_{G,0}$, feasibility cuts \eqref{FEAS_CUTS} are added. The optimality subproblem for vertex $v$ and fixed active generator power dispatch $\hat{P}_{G,0}^{i,(J)}$ from the master problem can be stated at iteration $J$ as:
\begin{align}
    &	\min_{P_{G,0}^i, W_v^{i}, \gamma_v} \, \,  \mu_v \gamma_v  \label{ObjSub1} \\
    	\text{s.t.\, } &  \eqref{PnodalRect} - \eqref{R_Conv_Up_lim_AP}, \,	\text{Tr} \{ \textbf{Y}^{i}_k W_v^{i} \} -  P_{{G_k},0}^i - P_{D_k}^i  = \nonumber \\ 
		 &	\quad \quad \sum_{w}^{n_W}  \zeta_{v_w} (d_{G_k}^{i} + \gamma_v d_{G_k}^{i} + d^{i}_{w_k}) \label{PG0} \\
    	& P_{G,0}^i= \hat{P}_{G,0}^{i,(J)} \quad: \Lambda^{v,(J)}
\end{align}
If the subproblem is feasible, then an optimality cut in form of \eqref{OPT_CUTS} is added to the master problem based on the Lagrangian multipliers $\Lambda^{v,(J)}$. If the subproblem is infeasible or the penalty weight $\mu_v$ is zero, we solve the following feasibility subproblem and add one feasibility cut \eqref{FEAS_CUTS} based on the Lagrangian multipliers $\Omega^{v,(J)}$ to the master problem:
\begin{align}
    &	\min_{P_{G,0}^i, W_v^{i}, \gamma_v, s\geq0} \, \, S_v= \sum_{i=0}^{n_{\text{grid}}} \sum_{k \in \mathcal{N}^{i}} \underline{s}^i_{P_k}+ \overline{s}^i_{P_k} + \underline{s}^i_{Q_k} +\overline{s}^i_{Q_k}  \label{ObjSub2}  \\
  & 	\text{s.t.\, }   \eqref{VnodalRect} - \eqref{LinkHVDC2},\, \eqref{PG0}  \\
    &	 \hphantom{\text{s.t.\, } }	 	\underline{P}^{i}_k -  \underline{s}^i_{P_k} \leq \text{Tr}\{ \textbf{Y}^{i}_k W_v^{i} \} + P^{i}_{D_k} \nonumber \\
    &	 \hphantom{\text{s.t.\, } }	 \quad \quad \quad \quad \quad \quad  - P_{W_k}^f - \overline{\zeta}_{v_k} \leq \overline{P}^{i}_k  + \overline{s}_{P_k}^i \\
&	\hphantom{\text{s.t.\, } }		\underline{Q}^{i}_k -  \underline{s}_{Q_k}^i  \leq \text{Tr}\{ \bar{\textbf{Y}}^{i}_{k} W_v^{i}  \}  + Q^{i}_{D_k} \leq \overline{Q}^{i}_k  + \overline{s}_{Q_k}^i    \\
  &  \hphantom{\text{s.t.\, } }		 P_{G,0}^i= \hat{P}_{G,0}^{i,(J)} \quad: \Omega^{v,(J)}
\end{align}
In the feasibility subproblem, we introduce the slack variables $\underline{s}_{P_k}$, $\overline{s}_{P_k}$, $\underline{s}_{Q_k}$, $\overline{s}_{Q_k}$ to allow for curtailment and over-satisfaction of active and reactive power at each bus, respectively. In the objective function \eqref{ObjSub2}  we minimize the sum of the slack variables $S_v$, i.e. the amount of constraint violation. The trivial solution of zero active and reactive power nodal injections is included, hence ensuring the existence of a solution to the feasibility subproblem. Using only a subset of possible slack variables had two desirable effects in our numerical studies: First, it reduced the computational time, as less slack variables need to be considered. Second, the amount of iterations was decreased.

The Benders algorithm includes the following steps. First, the master problem is solved without optimality and feasibility cuts. With the obtained generation dispatch $\hat{P}_{G,0}^{i,(J)}$, one optimality subproblem is solved for each vertex of the uncertainty set. If the subproblem is infeasible or the penalty term $\mu_v$ is zero, then the feasibility subproblem is solved and a feasibility cut is added if the sum of the slack variables $S_v$ is larger than zero. The algorithm continues to iteratively solve master and subproblems and terminates when the difference in the value of auxiliary variable $\Theta$ and the sum of the objective values of the subproblems $\sum_{v \in \mathcal{V}} \mu_v \gamma_v$ is lower than a specified tolerance, and the sum of the slack variables in the feasibility subproblems $\sum_{v \in \mathcal{V}} S_v$ is lower than a specified tolerance.

The advantage of applying Benders decomposition is that it allows us to deal with high dimensions of uncertainty, (i.e. a large number of vertices), since the complexity of one subproblem remains constant independent of the number of uncertain injections. Assuming that the subproblems are solved in parallel, the complexity of the subproblem is comparable to the complexity of solving the OPF without considering uncertainty. Note that this framework also allows to include security constraints in a straightforward way as e.g. in \cite{Wang2008} for the security constrained unit commitment problem, which would result in one subproblem for each vertex and each outage. Last but not least, a significant additional benefit of this framework is that, contrary to non-convex AC-OPF formulations, the convex SDP formulation guarantees theoretical convergence of the Benders decomposition \cite{birge2011introduction}.

\section{Simulation And Results}\label{IV}
\begin{figure}
\center
 \resizebox{!}{5.25cm}{
     \input{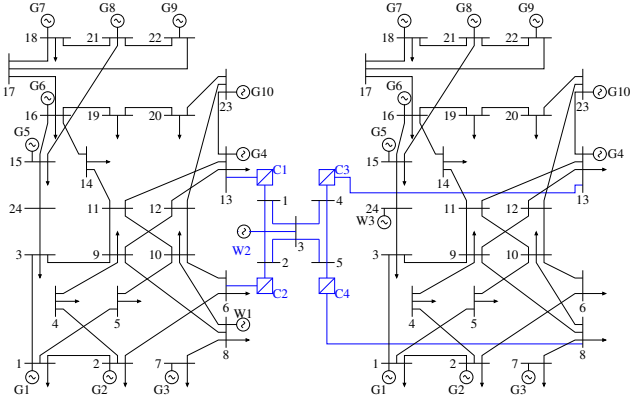}
     }
     \vspace{-4pt}
    \caption{Two IEEE 24 bus systems interconnected with a 5 bus multi-terminal HVDC grid and two onshore and one offshore wind farm.}
    \label{24bus}
\end{figure}
\begin{table}
\caption{Simulation parameter}
\footnotesize
\center
\begin{tabular}{l c}
\toprule
Confidence interval $1-\epsilon$ and parameter $\beta$ & 0.95 $\vert$  $10^{-3}$ \\
Wind farm reactive limits on $\tau$ ($\cos \phi = 0.95$) & $\pm 0.3827$ \\
\midrule
HVDC line resistance (p.\,u.\,) & 0.01  \\
HVDC upper and lower voltage limits $\overline{V}$, $\underline{V}$ (p.u.) & 1.1 $\vert$ 0.9 \\
Converter apparent power $S_C^{\text{nom}}$ (MVA) & 200 \\
Converter voltage rating (kV) & 240 \\
\midrule
Resistance $R_{T_k}$, reactance $X_{T_k}$ (p.u.) \cite{beerten2012generalized}  & 0.0015 $\vert$  0.1121\\
Converter resistance $R_{C_k}$, reactance $X_{C_k}$ (p.u.) \cite{beerten2012generalized}  & 0.0001 $\vert$  0.1643  \\
Converter loss terms $a_k$, $c_k$ (p.u.) \cite{beerten2012generalized} & 0.0110 $\vert$  0.0069 \\
Filter $B_f$ (multi-modular converter has no filter) & -- \\
\midrule
Upper converter current limit & $\tfrac{1}{1.1} \cdot S_C^{\text{nom}}$ \\
Voltage modulation $m$ & 1.05\\
Upper and lower converter reactive limits $m_c$, $m_b$ & 0.4 $\vert$  0.5 \\
\bottomrule
\end{tabular}
\label{SIM_PARAM}
\end{table}
The optimization problem is implemented in MATLAB using the optimization toolbox YALMIP \cite{Lofberg2004} and the SDP solver MOSEK 8 \cite{mosek}. A small resistance of $10^{-4} \, \text{p.u.}$ has to be added to each transformer, which is a condition for obtaining an exact solution , i.e. rank-1 solution matrices \cite{lavaei2012zero}. We do not include the slack bus angle constraint. Therefore, to investigate whether the rank of an obtained solution matrix $W^{i}$ is 2, the ratio $\rho$ of the $2^{\text{nd}}$ to $3^{\text{rd}}$ eigenvalue of each maximum clique $clq$ is computed, a measure proposed by \cite{molzahn2013implementation}. This value should be around $10^5$ or larger to imply that the obtained solution matrix is rank-2. The respective rank-1 solution can be retrieved by following the procedure in \cite{molzahn2013implementation}. The obtained solution is then a feasible solution to the original non-convex AC-OPF. The work in \cite{madani2015convex} proposes the use of the following measure to evaluate the degree of the near-global optimality of a penalized relaxation: Let $\tilde{f}_1(x)$ be the generation cost of the convex OPF without a penalty term and $\tilde{f}_2(x)$ the generation cost with a penalty weight sufficiently high to obtain rank-1 matrices, i.e. it corresponds to a solution that is feasible to the non-convex chance constrained AC-OPF using the piecewise affine approximation. The near-global optimality can be assessed by computing the parameter $\delta_\text{opt}:= \tfrac{\tilde{f}_1(x)}{\tilde{f}_2(x)} \cdot 100 \%$. Note that this distance is an upper bound to the distance from the global optimum. 

For the following analysis, we consider a 53-bus AC-DC system, i.e. two IEEE 24 bus systems interconnected with a 5 bus HVDC grid shown in Fig.~\ref{24bus}. A total of three on- and offshore wind farms with a rated power of 150~MW, 300~MW and 400~MW are placed at bus 8 of the first AC grid, at bus 24 of the second AC grid, and at bus 3 of the HVDC grid, respectively. Table~\ref{SIM_PARAM} shows the simulation parameters. For the generator participation factors, each generator adjusts its active power proportional to its maximum active power. For the first two subsections, we solve the proposed OPF formulation as one optimization problem. In Section~\ref{sub_benders}, we replace the IEEE 24 bus systems with IEEE 118 bus systems, i.e. we consider a 241-bus AC-DC system, and solve the decomposed OPF formulation with one subproblem for each vertex of the uncertainty set as proposed in Section~\ref{III.6}.
\subsection{Systematic Procedure to Obtain Rank-1 Solution Matrices}
In this section, we showcase the systematic, heuristic procedure to identify rank-1 solution matrices. For illustrative purposes, we assume for each wind farm a forecasted infeed of 50\% of rated power and assume the forecast error bounds are within $\pm$25\% of rated power with 95\% probability. We select the penalty weight step size $\Delta \mu$ to be 25. 
\begin{figure}
	\begin{flushright}
		\begin{footnotesize}
			\input{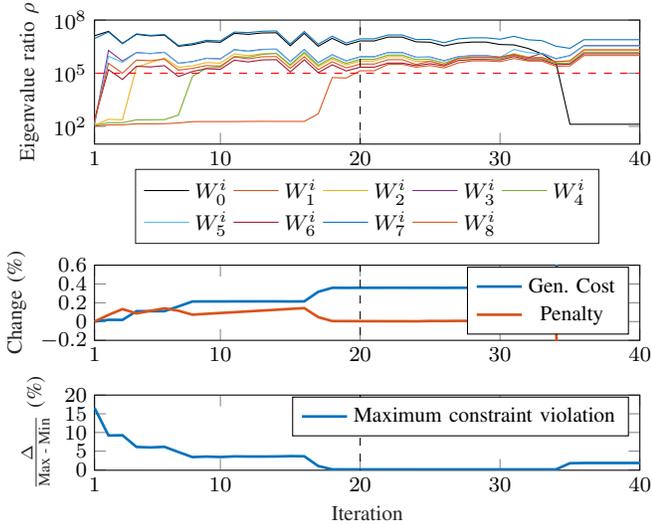} \quad \\
%
%
\definecolor{mycolor1}{rgb}{0.00000,0.44700,0.74100}%
\definecolor{mycolor2}{rgb}{0.85000,0.32500,0.09800}%
\begin{tikzpicture}

\begin{axis}[%
width=7.25cm,
height=1.0cm,
scale only axis,
xmin=1,
xmax=40,
xtick = {1,10,20,30,40,50},
xlabel style={font=\color{white!15!black}},
ymin=-0.2,
ymax=0.6,
ylabel style={font=\color{white!15!black}},
ylabel={Change (\%)},
axis background/.style={fill=white},
]
\addplot [color=mycolor1,line width=1pt]
  table[row sep=crcr]{%
1	0\\
2	0.0184599541365884\\
3	0.0174383138417369\\
4	0.110172264512286\\
5	0.110148258323321\\
6	0.110238187056538\\
7	0.160878532825024\\
8	0.213199501786093\\
9	0.213884375042866\\
10	0.213836601095778\\
11	0.214545939050808\\
12	0.214507381272242\\
13	0.214576740604116\\
14	0.214591800778237\\
15	0.213418606827219\\
16	0.214590223354023\\
17	0.316465247859867\\
18	0.359025294325562\\
19	0.358314932088732\\
20	0.359064427608402\\
21	0.359064427608402\\
22	0.359444754433923\\
23	0.359444754433923\\
24	0.359062994331367\\
25	0.35909856511411\\
26	0.358631479786681\\
27	0.359165512507118\\
28	0.359274260532217\\
29	0.359274260532217\\
30	0.359147902739139\\
31	0.359418119066319\\
32	0.35913265159266\\
33	0.359160743952629\\
34	0.358575528239172\\
35	5.17718882500664\\
36	5.17873861799836\\
37	5.17873861799836\\
38	5.17873861799836\\
39	5.17873861799836\\
40	5.17873861799836\\
41	5.17873861799836\\
42	5.17873861799836\\
43	5.17873861799836\\
44	5.17873861799836\\
45	5.17873861799836\\
46	5.17873861799836\\
47	5.17873861799836\\
48	5.17873861799836\\
49	5.17873861799836\\
50	5.17873861799836\\
};
\addlegendentry{Gen. Cost}

\addplot [color=mycolor2,line width=1pt]
  table[row sep=crcr]{%
1	0\\
2	0.0706934341008096\\
3	0.132462746078654\\
4	0.086640964776854\\
5	0.113589869189231\\
6	0.140529159415886\\
7	0.115311895239928\\
8	0.0730540745942781\\
9	0.0823072282800715\\
10	0.0910011244975503\\
11	0.0994723120513641\\
12	0.108178504591861\\
13	0.116845158156427\\
14	0.125530405394995\\
15	0.134515308557163\\
16	0.142904884993049\\
17	0.0470358490985351\\
18	0.00493946148342787\\
19	0.0046696050194485\\
20	0.0038172847975789\\
21	0.0038172847975789\\
22	0.00317613680706825\\
23	0.00317613680706825\\
24	0.00264895137199947\\
25	0.00585116977173659\\
26	0.00535170136500855\\
27	0.00691455932179077\\
28	0.0064071195156824\\
29	0.0064071195156824\\
30	0.00732232850604141\\
31	0.00709421605081087\\
32	0.00669214438799074\\
33	0.013962795208292\\
34	0.0155845943995053\\
35	-5.34904388604742\\
36	-5.60660880597889\\
37	-5.60660880597889\\
38	-5.60660880597889\\
39	-5.60660880597889\\
40	-5.60660880597889\\
41	-5.60660880597889\\
42	-5.60660880597889\\
43	-5.60660880597889\\
44	-5.60660880597889\\
45	-5.60660880597889\\
46	-5.60660880597889\\
47	-5.60660880597889\\
48	-5.60660880597889\\
49	-5.60660880597889\\
50	-5.60660880597889\\
};
\addlegendentry{Penalty}

\addplot [color=black,dashed]
  table[row sep=crcr]{%
20.0	-2.0\\
20.0000001	2.0\\
};
\end{axis}
\end{tikzpicture}%
 \quad \\
%
%
\definecolor{mycolor1}{rgb}{0.00000,0.44700,0.74100}%
\definecolor{mycolor2}{rgb}{0.85000,0.32500,0.09800}%
\begin{tikzpicture}

\begin{axis}[%
width=7.25cm,
height=1.0cm,
scale only axis,
xmin=1,
xmax=40,
xtick = {1,10,20,30,40,50},
xlabel style={font=\color{white!15!black}},
xlabel={Iteration},
ymin=-0.1,
ymax=20,
ylabel style={font=\color{white!15!black}},
ylabel={$\tfrac{\Delta}{\text{Max - Min}}$ (\%)},
axis background/.style={fill=white},
]
\addplot [color=mycolor1,line width=1pt]
  table[row sep=crcr]{%
1	16.5747880134243\\
2	9.18983202433028\\
3	9.25481057893949\\
4	6.13344825517829\\
5	5.99458805042104\\
6	6.16039046675421\\
7	4.77006672031551\\
8	3.42012137330134\\
9	3.5563765541003\\
10	3.43674180733268\\
11	3.60793498133902\\
12	3.56745826609338\\
13	3.56339396937936\\
14	3.59238540942647\\
15	3.67777180007759\\
16	3.63507569662012\\
17	1.0257281953397\\
18	0.13310492794966\\
19	0.131405075502533\\
20	0.127729052583995\\
21	0.127729052583995\\
22	0.12681529306254\\
23	0.12681529306254\\
24	0.126309131916669\\
25	0.125896724818723\\
26	0.12504953443541\\
27	0.12553195527684\\
28	0.125601061902803\\
29	0.125601061902803\\
30	0.125235325791036\\
31	0.125663032203405\\
32	0.124884169888576\\
33	0.124602770955902\\
34	0.122020878317364\\
35	1.77992165485669\\
36	1.84564972402117\\
37	1.84564972402117\\
38	1.84564972402117\\
39	1.84564972402117\\
40	1.84564972402117\\
41	1.84564972402117\\
42	1.84564972402117\\
43	1.84564972402117\\
44	1.84564972402117\\
45	1.84564972402117\\
46	1.84564972402117\\
47	1.84564972402117\\
48	1.84564972402117\\
49	1.84564972402117\\
50	1.84564972402117\\
};
\addlegendentry{Maximum constraint violation}

\addplot [color=black,dashed]
  table[row sep=crcr]{%
20.0	-2.0\\
20.0000001	22.0\\
};
\end{axis}
\end{tikzpicture}%
 \quad \\
		\end{footnotesize}
		\vspace{-12pt}
	\end{flushright}
	\caption{Minimum eigenvalue ratios $\rho$ over all grids $i$, generation cost and penalty for each iteration of the systematic procedure for the considered test case. The change of generation cost and penalty is normalized to the non-penalized objective value. We report the maximum constraint violation at each iteration evaluated with non-linear AC power flows using the obtained set-points for the forecasted system state and the vertices. The procedure would terminate at iteration 20. In this test case only, we extend the number of iterations to 40 to further investigate the relationship between eigenvalue ratio and penalty parameter. For this purpose, once we reach the defined minimum clique eigenvalue ratio, we double it.}
	\label{EV_ratio}
\end{figure}
In Fig.~\ref{EV_ratio}, we show minimum eigenvalue ratios $\rho$ of all grids $i$, generation cost and penalty for each iteration of the proposed systematic procedure using the active power loss penalty in \eqref{ObjPen}. Furthermore, we report the maximum constraint violation at each iteration evaluated with non-linear AC power flows using the obtained set-points for the forecasted system state and the vertices of the uncertainty set. For this purpose we divide the maximum occurring constraint violation $\Delta$ by the difference in maximum and minimum constraint limit (Max - Min). At iteration 20 we obtain rank-2 matrices $W_{0-8}^{i}$ and the corresponding rank-1 solution matrices can be recovered following \cite{molzahn2013implementation} by means of an eigendecompostion. At this point, the near-global optimality guarantee evaluates to 99.52\%, i.e. the distance to the global optimum is at most 0.48\% and the obtained set-points comply with the constraints for the forecasted system state and the vertices. The proposed procedure allows for a systematic identification of suitable penalty weights. This improves upon previous works \cite{venzke2017, madani2016promises} which use an ad-hoc defined penalty parameter. In case simulations with similar setup are rerun, the previously obtained penalty weights $\mu_v$ can be used as hot start. The active loss power penalty effectively minimizes the deviation in active power between the vertices of the uncertainty set and the forecasted system state. Note that if we increase the penalty weights to a value too high, we obtain a higher rank solution for the forecasted system $W_0$ state (here in iteration 35), which corresponds to a non-physical solution, and this elucidates the importance of a systematic method to choose the penalty weights to obtain rank-1 solution matrices.
\subsection{Monte Carlo Analysis Using Realistic Forecast Data}
\label{IV.2}
\begin{table}
		\caption{Comparison of performance of AC-OPF without considering uncertainty, the chance-constrained AC-OPF formulation and a chance-constrained DC-OPF \cite{wiget2014probabilistic} using 10'000 samples from realistic forecast data. Insecure instances are marked bold. The CC-AC-OPF uses the active loss penalty \eqref{ObjPen}.}
	\label{SampledGauss}
	\centering
    \begin{tabular}{l c c c c c}
    \toprule
    Time step (h) & 1 & 2 & 3 & 4 & 5 \\
    \midrule
    \multicolumn{6}{c}{Empirical bus voltage constraint violation probability (\%) }\\
   \midrule
    AC-OPF w/o CC \quad \quad \quad  & \textbf{24.07}  & \textbf{24.24} &  \textbf{15.71} &  \textbf{13.85} &  \textbf{23.35}  \\
   \midrule
  Penalized CC-AC-OPF & 0.0 & 0.0 & 0.0& 0.0 & 0.0  \\
   \midrule
   CC-DC-OPF \cite{wiget2014probabilistic}  & \textbf{100.0} & \textbf{100.0} & \textbf{100.0}& \textbf{100.0} & \textbf{100.0}  \\
   \midrule
    \multicolumn{6}{c}{Empirical generator active power constraint violation probability (\%)}\\
   \midrule
    AC-OPF w/o CC  & \textbf{100.0} & \textbf{100.0}& \textbf{100.0}&\textbf{100.0}  &\textbf{100.0}  \\
   \midrule
   Penalized CC-AC-OPF  & 0.0  & 0.0 & 0.0& 0.0 & 0.0 \\
   \midrule 
  CC-DC-OPF \cite{wiget2014probabilistic} & \textbf{9.95} &  \textbf{2.45}  & \textbf{11.16}  &  \textbf{1.19}  & \textbf{ 0.83}\\
   \midrule

    \multicolumn{6}{c}{Empirical apparent branch flow constraint violation probability (\%) }\\
   \midrule
    AC-OPF w/o CC  & 0.0 & 0.0 & 0.0 &0.0 & 0.0 \\
   \midrule
   Penalized CC-AC-OPF & 0.0  & 0.0& 0.0& 0.0 &0.0   \\
   \midrule
   CC-DC-OPF \cite{wiget2014probabilistic}  & 0.0& 0.0&0.0 &0.0 & 0.0\\
   \midrule
       \multicolumn{6}{c}{Cost of uncertainty (\%)}\\
    \midrule
  Penalized CC-AC-OPF  & 3.94 &  4.78 & 3.90 & 5.02 & 5.46   \\
   \midrule 
  CC-DC-OPF \cite{wiget2014probabilistic}  & 1.36 &   2.26  &  1.87 &   3.06 & 3.51 \\
   \bottomrule
    \end{tabular}
    \label{ViolProb}
\end{table}

In this section, we compare the proposed chance constrained AC-OPF using the active power loss penalty in \eqref{ObjPen} to a DC-OPF formulation \cite{wiget2014probabilistic} and to an AC-OPF without considering uncertainty, for the test case shown in Fig.~\ref{24bus} using realistic forecast data. Note that in literature the application of chance constraints to interconnected AC and HVDC grids is limited to the DC-OPF formulation. We select the penalty weight step size $\Delta \mu$ to be 100. The DC-OPF includes a joint chance constraint on active generator power and active line flows, and corrective control of active power set-points of HVDC converters. As the branch flow limits are specified in terms of apparent power, for the DC-OPF only we set the maximum active branch flow to 80\% $\overline{S}_ {lm}$. To construct the rectangular uncertainty set for both formulations, we draw $N_S = 377$ samples according to \eqref{NumberOfSamples} with $\epsilon = 0.05$ and $\beta = 10^{-3}$. The sample base representing realistic wind day-ahead forecast data has been constructed exactly following the procedure in \cite{pinson2013wind} and is based on wind power measurements in the Western Denmark area from 15 different control zones collected by the Danish transmission system operator Energinet. We select control zone 1, 11, 3 to correspond to the wind farm at bus 8, 24, and 3, respectively. We use three different sets of $N_s$ samples to run the following computational experiments and report the averaged results. Note that we use the same sample base for both drawing the $N_s$ samples and the Monte Carlo analysis.  The converter C2 is selected as DC slack bus which compensates the possible mismatch between set-points and the realized active power flows. Note that the DC-OPF approach does not model converter losses and that the AC-OPF without considering uncertainty includes no corrective control policies, i.e. resulting mismatches are compensated via the slack bus converter. With our approach we  include suitable HVDC converter corrective control policies. 

In order to evaluate the empirical constraint violation probabilities of the three approaches, we run a Monte Carlo analysis using AC-DC power flows of MATACDC \cite{beerten2015development} with 10'000 samples drawn from the realistic forecast data sample base. MATACDC is a sequential AC/DC power flow solver interfaced with MATPOWER \cite{zimmerman2011matpower} which uses the HVDC converter model shown in Section~\ref{II.1}. The DC-OPF provides only the active power set-points for generators and HVDC converters. To exclude numerical errors, a minimum violation limit of $10^{-3}$ per unit for generator limits on active power and 0.1\% for voltage and apparent line flow limits is assumed. In the AC power flow the generator reactive power limits are enforced to avoid a possibly high non-physical overloading of the limits. Furthermore, we distribute the loss mismatch from the active generator set-points among the generators according to their participation factors and rerun the power flow to mimic the response of automatic generation control (AGC). 

\begin{figure}
	\begin{footnotesize}
	\input{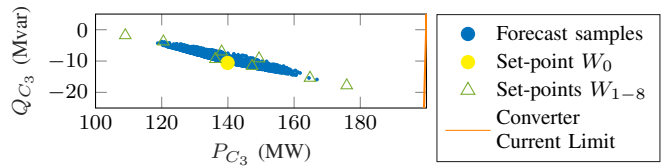}
	\end{footnotesize}
	\caption{Corrective control policy of converter active and reactive power set-points and 10'000 sample realizations for converter C3 and time step 3.}
	\label{HVDC_CORRFig}
\end{figure} 
\begin{table}
\center
		\caption{Comparison of active power loss and reactive power penalty }
		 \begin{tabular}{l c c c c c}
		   \toprule
          \multicolumn{6}{c}{Near-global optimality guarantee $\delta_\text{opt}$ (\%)}\\
              \midrule
          		       Time step (h) & 1 & 2 & 3 & 4 & 5 \\
    \midrule 
   Active loss penalty  &  99.64 & 99.61 & 99.53 & 99.63 & 99.58  \\
      Reactive power penalty &  96.57 & 96.33 & 91.94 & 95.82 & 94.15  \\
   \bottomrule
    \end{tabular}
    \label{NearGlobal}
\end{table}
In Table~\ref{ViolProb} the resulting violation probability of the joint chance constraint on active power, bus voltages, and active branch flows, and the cost of uncertainty are compared for our approach, an AC-OPF without considering uncertainty and the chance constrained DC-OPF. We find that our proposed approach complies with the joint chance constraint. If we do not consider uncertainty in the AC-OPF, violations of generator and voltage limits occur. The chance constrained DC-OPF violates the target value of $\epsilon = 5\%$ as well. Violations of voltage limits occur as the DC-OPF approximation does not model voltage magnitudes. As losses which can make up several percent of load are also neglected, the limits on generator active power are violated as well. The cost of uncertainty, i.e. the additional cost incurred by taking uncertainty into account, is lower for the DC-OPF approach as the cost for the active power losses are not included. On average, for our approach, using a laptop with Intel \mbox{i7-7820HQ} CPU @ 2.90 GHz and 32 GB RAM, the total solving time is 13.0 seconds, the individual SDP solving time is 1.9 seconds and the number of iterations for the systematic procedure to obtain rank-1 solution matrices is 6.8. For the HVDC converter C3 and time step 3, we show in Fig.~\ref{HVDC_CORRFig} the active and reactive power set-points from our approach for the first sample set and the resulting 10'000 realizations which comply with the HVDC converter limits.

In Table~\ref{NearGlobal} we compare the near-global optimality guarantee that we obtain by using the active loss penalty from \eqref{ObjPen} and the reactive power penalty \eqref{ObjPenQ} averaged for the three $N_s$ sample sets. For both approaches, we us0e the same penalty weight step size $\Delta \mu = 100$ for the systematic procedure. By using the active loss penalty, the near-global optimality guarantee evaluates to at least 99.5\% for all considered time steps, i.e. the distance to the global optimum is at most 0.5\%. The upper bound on the sub-optimality incurred by the reactive power loss penalty is substantially larger with an average of 5.1\% which is more than ten times larger than the bound obtained by the active power loss penalty, i.e. the obtained generation cost is higher for the reactive power penalty by that amount. This highlights the effectiveness of the proposed active power loss penalty to obtain rank-1 solution matrices without incurring a significant sub-optimality.

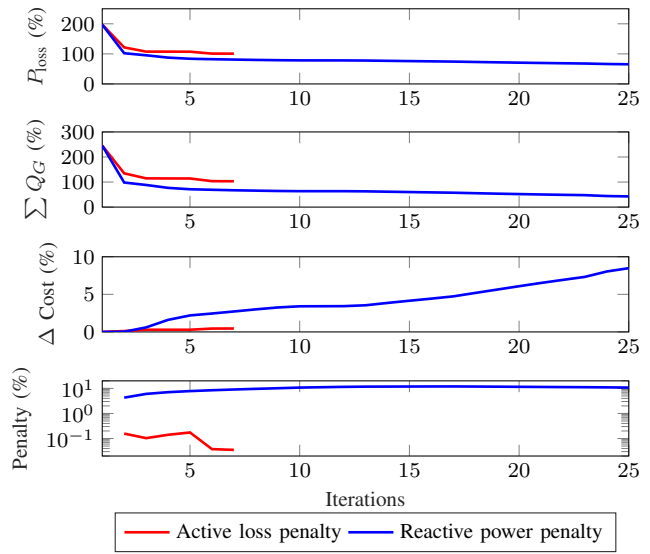
\begin{figure}
	\begin{flushright}
		\begin{footnotesize}
%
%
\definecolor{mycolor1}{rgb}{0.00000,0.44700,0.74100}%
\definecolor{mycolor2}{rgb}{0.85000,0.32500,0.09800}%
\begin{tikzpicture}

\begin{axis}[%
width=7cm,
height=1.0cm,
scale only axis,
xmin=1,
xmax=25,
xlabel style={font=\color{white!15!black}},
ymin=0,
ymax=250,
ylabel style={font=\color{white!15!black}},
ylabel={$P_{\text{loss}}$ (\%)},
axis background/.style={fill=white},
legend style={legend cell align=left, align=left, draw=white!15!black}
]
\addplot [color=red,line width=1pt]
  table[row sep=crcr]{%
1	197.11468920102\\
2	121.486620227627\\
3	107.510182330326\\
4	107.404552548437\\
5	107.230406151269\\
6	100.838376917514\\
7	100.766053823608\\
};

\addplot [color=blue,line width=1pt]
  table[row sep=crcr]{%
1	197.11468920102\\
2	102.126869932625\\
3	95.3085150926878\\
4	87.5404438354079\\
5	83.765368657455\\
6	82.2827452323855\\
7	81.0161394693769\\
8	79.8979863728065\\
9	78.9454150964942\\
10	78.3772981614708\\
11	78.307829926535\\
12	78.2431197898824\\
13	77.8215522819839\\
14	76.8661261467017\\
15	75.9792166266986\\
16	75.1513075254082\\
17	74.3043660309847\\
18	73.0415667466027\\
19	71.863461611663\\
20	70.7662441475391\\
21	69.7394465380077\\
22	68.7868752616954\\
23	67.8990141220357\\
24	66.2983898595409\\
25	65.3819801301816\\
};

\end{axis}
\end{tikzpicture}%
 \\
%
%
\definecolor{mycolor1}{rgb}{0.00000,0.44700,0.74100}%
\definecolor{mycolor2}{rgb}{0.85000,0.32500,0.09800}%
\begin{tikzpicture}

\begin{axis}[%
width=7cm,
height=1.0cm,
scale only axis,
xmin=1,
xmax=25,
xlabel style={font=\color{white!15!black}},
ymin=0,
ymax=300,
ylabel style={font=\color{white!15!black}},
ylabel={$\sum Q_{G}$ (\%)},
axis background/.style={fill=white},
legend style={legend cell align=left, align=left, draw=white!15!black}
]
\addplot [color=red,line width=1pt]
  table[row sep=crcr]{%
1	245.488581568293\\
2	134.835371300089\\
3	114.692354287944\\
4	114.342426326903\\
5	114.221106559481\\
6	103.579532037659\\
7	103.128133523612\\
};

\addplot [color=blue,line width=1pt]
  table[row sep=crcr]{%
1	245.488581568293\\
2	98.0462212913625\\
3	88.3325975434886\\
4	76.7599921972053\\
5	71.2646687405673\\
6	69.0973398630404\\
7	67.1930130765256\\
8	65.6863550340003\\
9	64.3360951119598\\
10	63.5763493187955\\
11	63.5361375482113\\
12	63.5400731683111\\
13	62.9191007621414\\
14	61.4324630479495\\
15	60.0517448486155\\
16	58.6992604456491\\
17	57.4049547745916\\
18	55.4477537876066\\
19	53.582783202089\\
20	51.9553187338939\\
21	50.3577992012402\\
22	48.9529539395557\\
23	47.6547126483985\\
24	44.1237769632755\\
25	42.6410748691834\\
};

\end{axis}
\end{tikzpicture}%
 \\
%
%
\definecolor{mycolor1}{rgb}{0.00000,0.44700,0.74100}%
\definecolor{mycolor2}{rgb}{0.85000,0.32500,0.09800}%
\begin{tikzpicture}

\begin{axis}[%
width=7cm,
height=1.0cm,
scale only axis,
xmin=1,
xmax=25,
xlabel style={font=\color{white!15!black}},
ytick={0,5,10},
ymin=0,
ymax=10,
ylabel style={font=\color{white!15!black}},
ylabel={$\Delta$ Cost (\%)},
axis background/.style={fill=white},
legend style={legend cell align=left, align=left, draw=white!15!black}
]
\addplot [color=red, line width=1pt]
  table[row sep=crcr]{%
1	0\\
2	0.116561437591045\\
3	0.291403593977634\\
4	0.291403593977634\\
5	0.291403593977634\\
6	0.44681884409907\\
7	0.456532297231661\\
};

\addplot [color=blue,line width=1pt]
  table[row sep=crcr]{%
1	0\\
2	0.0291403593977577\\
3	0.602234094220492\\
4	1.60271976687712\\
5	2.19524040796502\\
6	2.44779018941233\\
7	2.71976687712481\\
8	3.00145701796988\\
9	3.25400679941718\\
10	3.40942204953861\\
11	3.4191355026712\\
12	3.42884895580377\\
13	3.54541039339485\\
14	3.85624089363768\\
15	4.14764448761534\\
16	4.4293346284604\\
17	4.73045167557065\\
18	5.18698397280231\\
19	5.63380281690138\\
20	6.08062166100048\\
21	6.51772705196696\\
22	6.92569208353568\\
23	7.32394366197182\\
24	8.04273919378338\\
25	8.47984458474986\\
};

\end{axis}
\end{tikzpicture}%
 \\
%
%
\definecolor{mycolor1}{rgb}{0.00000,0.44700,0.74100}%
\definecolor{mycolor2}{rgb}{0.85000,0.32500,0.09800}%
\begin{tikzpicture}

\begin{axis}[%
width=7cm,
height=1.0cm,
scale only axis,
xmin=1,
xmax=25,
xlabel style={font=\color{white!15!black}},
xlabel={Iterations},
ymin=0.02,
ymax=20,
ymode=log,
ylabel style={font=\color{white!15!black}},
ylabel={Penalty (\%)},
axis background/.style={fill=white},
legend style={legend style={at={(0.5,-0.75)},anchor=north},legend cell align=left, align=left, draw=white!15!black, legend columns=2}
]
\addplot [color=red,line width=1pt]
  table[row sep=crcr]{%
1	0\\
2	0.157304225352113\\
3	0.103395046138902\\
4	0.14060660514813\\
5	0.174094317629917\\
6	0.0377348227294803\\
7	0.0352145701796989\\
};
\addlegendentry{Active loss penalty}

\addplot [color=blue,line width=1pt]
  table[row sep=crcr]{%
1	0\\
2	4.27489072365226\\
3	6.03011170471102\\
4	7.07139388052452\\
5	7.80767362797474\\
6	8.44099077221952\\
7	9.04613890237979\\
8	9.59689169499757\\
9	10.1457017969888\\
10	10.7809616318601\\
11	11.1452161243322\\
12	11.5075279261778\\
13	11.7552209810588\\
14	11.8144730451676\\
15	11.8737251092763\\
16	11.9368625546382\\
17	11.9708596406022\\
18	11.8416707139388\\
19	11.7095677513356\\
20	11.5590092277805\\
21	11.4181641573579\\
22	11.2812044681884\\
23	11.1491015055852\\
24	10.9917435648373\\
25	10.7887323943662\\
};
\addlegendentry{Reactive power penalty}

\end{axis}
\end{tikzpicture}%
 \\
		\end{footnotesize}
	\end{flushright}
	\caption{The average active power loss $P_{\text{loss}}$, the average sum of generator reactive power $\sum Q_G$, the change in generation cost and the penalty term for time step 3 for the active power loss and reactive power generator penalty terms. The iterations are shown until rank-1 solution matrices are obtained. Note that all quantities are normalized by the corresponding values of the forecasted system state $W_0$ for the non-penalized CC-AC-OPF and the first two are averaged over all vertices of the uncertainty set.}
	\label{Pen_Comparison}
\end{figure}
To provide more insight into the active power loss and reactive power generator penalty, Fig.~\ref{Pen_Comparison} shows the average active power loss $P_{\text{loss}}$, the average sum of generator reactive power $\sum Q_G$, the change in generation cost and the penalty term for time step 3 of Table~\ref{NearGlobal}. We consider the first $N_s$ sample set. The iterations are shown until rank-1 solution matrices are obtained. Note that all quantities are normalized by the corresponding values of the forecasted system state $W_0$ of the non-penalized chance constrained AC-OPF and the first two are averaged over all vertices of the uncertainty set. We observe that for the non-penalized formulation, i.e. at iteration 1, both the high average active power loss of 197.1\% and the high average generator reactive power of 245.5\% indicate that several of the solution matrices $W_v$ correspond to non-physical higher rank solutions. Furthermore, the active power loss and the generator reactive power are coupled, and both heuristic penalty terms reduce these values from high non-physical values until a rank-1 solution can be recovered. However, in this case, the active power loss penalty requires significantly less iterations, 7 compared to 25, and a two orders of magnitude smaller penalty term to recover rank-1 solution matrices. An intuitive explanation is that the active power loss penalty penalizes directly the change in active power losses with respect to the forecasted system state, represented by $\gamma_v$, and that this term is significantly smaller than the absolute sum of generator reactive power.

\subsection{Tuning of Confidence Parameter $\beta$}
In the previous section, we have shown that our proposed tractable chance constrained AC-OPF formulation achieves compliance with the maximum allowable joint chance constraint violation probability of $\epsilon = 5\%$. However, the actual observed empirical joint violation probability is $\epsilon^{\text{emp}} = 0\%$ in Table~\ref{ViolProb}. The underlying reason is that the methodology we employ to achieve a tractable formulation of the chance constraints does not make any assumption on the distribution of the forecast errors, i.e. we are robust against the worst-case distribution for defined violation probability $\epsilon$. In this section, we propose a procedure to adjust the confidence parameter $\beta$ to match the empirical violation probability  $\epsilon^{\text{emp}}$ with the maximum allowable  constraint violation probability $\epsilon$ more closely while systematically reducing the cost of uncertainty.

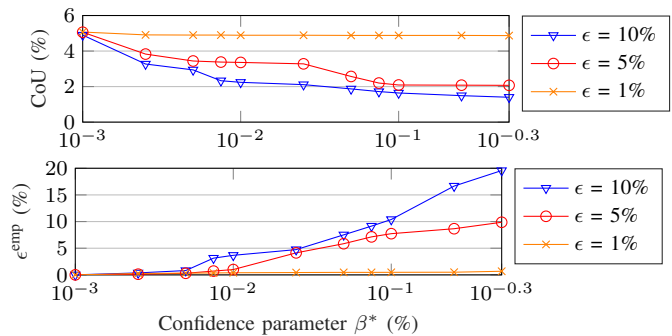
\begin{figure}
	\begin{flushright}
		\begin{footnotesize}
%
%
\begin{tikzpicture}

\begin{axis}[%
width=7.25cm,
height=3cm,
xmode=log,
xmin=0.001,
xmax=0.5,
xminorticks=true,
xlabel style={font=\color{white!15!black}},
xtick = {0.001,0.01,0.1,0.5},
ymin=0  ,
ymax=6,
ylabel style={font=\color{white!15!black}},
ylabel={CoU (\%)},
ymajorgrids,
axis background/.style={fill=white},
legend style={legend pos=outer north east,legend cell align=left, align=left, draw=white!15!black}
]
\addplot [color=blue, mark=triangle, mark options={solid, rotate=180, blue}]
  table[row sep=crcr]{%
0.001	4.90\\
0.0025	3.27\\
0.005	2.94\\
0.0075	2.33\\
0.01	2.24\\
0.025	2.11\\
0.05	1.87\\
0.075	1.72\\
0.1	    1.64\\
0.25	1.49\\
0.5	    1.40\\
};
\addlegendentry{$\epsilon$ = 10\%}

\addplot [color=red, mark=o, mark options={solid, red}]
  table[row sep=crcr]{%
0.001	5.06\\
0.0025	3.83\\
0.005	3.44\\
0.0075	3.38\\
0.01	3.36\\
0.025	3.28\\
0.05	2.57\\
0.075	2.20\\
0.1	    2.09\\
0.25	2.08\\
0.5	    2.07\\
};
\addlegendentry{$\epsilon$ = 5\%}

\addplot [color=orange, mark=x, mark options={solid, orange}]
  table[row sep=crcr]{%
0.001	 5.07 \\
0.0025	4.91\\
0.005	4.90\\
0.0075	4.90\\
0.01	4.89\\
0.025	4.89\\
0.05	4.88\\
0.075	4.88\\
0.1	    4.88\\
0.25	4.88\\
0.5	    4.87\\
};
\addlegendentry{$\epsilon$ = 1\%}

\end{axis}
\end{tikzpicture}%
 \quad \\
%
%
\begin{tikzpicture}

\begin{axis}[%
width=7.25cm,
height=3cm,
xmode=log,
xmin=0.001,
xmax=0.5,
xminorticks=true,
xlabel style={font=\color{white!15!black}},
xlabel={Confidence parameter $\beta^*$ (\%)},
ymin=0,
ymax=20,
ymajorgrids,
xtick = {0.001,0.01,0.1,0.5},
ylabel style={font=\color{white!15!black}},
ylabel={$\epsilon^{\text{emp}}$ (\%)},
axis background/.style={fill=white},
legend style={legend pos=outer north east,legend cell align=left, align=left, draw=white!15!black}
]
\addplot [color=blue, mark=triangle, mark options={solid, rotate=180, blue}]
  table[row sep=crcr]{%
0.001	0.0100\\
0.0025	0.3700\\
0.005	0.8100 \\
0.0075	3.1400    \\
0.01	3.6700\\
0.025	4.7400\\
0.05	7.5100\\
0.075	9.1500\\
0.1	    10.3900\\
0.25	16.6600\\
0.5	    19.5700\\
};
\addlegendentry{$\epsilon$ = 10\%}

\addplot [color=red, mark=o, mark options={solid, red}]
  table[row sep=crcr]{%
0.001	0\\
0.0025	0.1300\\
0.005	0.3000\\
0.0075	0.7100\\
0.01	1.0000\\
0.025	4.1300\\
0.05	5.8500\\
0.075	7.1300\\
0.1	    7.7300\\
0.25	8.6600\\
0.5	    9.8700\\
};
\addlegendentry{$\epsilon$ = 5\%}

\addplot [color=orange, mark=x, mark options={solid, orange}]
  table[row sep=crcr]{%
0.001	0\\
0.0025  0.2600\\
0.005	0.3300\\
0.0075	0.3600\\
0.01	0.4000 \\
0.025	0.4400\\
0.05	0.4700\\
0.075   0.4700\\
0.1	    0.4800\\
0.25	0.5000\\
0.5	    0.6800\\
};
\addlegendentry{$\epsilon$ = 1\%}

\end{axis}
\end{tikzpicture}%
 \quad 
		\end{footnotesize}
	\end{flushright}
	\caption{The cost of uncertainty (CoU) and the empirical joint violation probability $\epsilon^{\text{emp}}$ are shown as a function of the adjusted confidence parameter $\beta^*$ for fixed values of $\epsilon$ for time step 4. Starting from a value $\beta = 10^{-3}$, the amount of samples forming the rectangular uncertainty set is reduced according to the new confidence parameter $\beta^*$ in order to match $\epsilon^{\text{emp}}$ with $\epsilon$ more closely while systematically reducing the cost of uncertainty.}
	\label{CoU_Emp}
\end{figure} 
For this purpose, after we obtain the empirical violation probability $\epsilon^{\text{emp}}$ as a result of the Monte Carlo Analysis for a given $\epsilon$ and $\beta$, we adjust the confidence parameter $\beta$. Based on the new confidence parameter, which we denote with $\beta^*$, we compute the reduced number of samples according to \eqref{NumberOfSamples}, and discard samples from the initial $N_s$ samples of the rectangular uncertainty set. In the discarding process, we select the worst-case samples, i.e. the samples which when removed reduce one of the dimensions of the rectangular set the most. We can tune the confidence parameter $\beta^*$ iteratively, until the empirical violation probability  $\epsilon^{\text{emp}}$ matches the maximum allowable constraint violation probability $\epsilon$. Note, that in case of a minimum confidence parameter $\beta^*$ (here $\beta^*$ = 0.5) the solution is still conservative, the number of considered samples could be further reduced by also adjusting the violation probability $\epsilon$ in \eqref{NumberOfSamples}. This proposed procedure is similar to distributionally robust optimization, e.g. \cite{esfahani2018data} in which both the distance metric around the empirical distribution and violation probability $\epsilon$ are varied over a wide range of values to achieve a target empirical violation probability  $\epsilon^{\text{emp}}$. 

In Fig.~\ref{CoU_Emp}, for the previous simulation setup, the time step 4 and a penalty step size of $\Delta \mu = 10$, the cost of uncertainty and the empirical joint violation probability $\epsilon^{\text{emp}}$ are shown as a function of the adjusted confidence parameter $\beta^*$ for a fixed value of $\epsilon$. We consider the first $N_s$ sample set. By tuning the confidence parameter $\beta^*$, and subsequently discarding samples from the rectangular uncertainty set, we can match the empirical joint violation probability $\epsilon^{\text{emp}}$ with the maximum allowable joint violation probability $\epsilon$ more closely at a lower cost of uncertainty, and reduce the conservativeness of our approach. For $\epsilon = 5\%$, we can achieve an empirical violation probability $\epsilon^{\text{emp}} = 4.13\%$ with $\beta^* =0.025$, while reducing the cost of uncertainty from 5.06\% to 3.29\%. Similarly for $\epsilon = 10\%$, an empirical violation probability $\epsilon^{\text{emp}} = 9.15\%$ with $\beta^* =0.075$ is achieved, while reducing the cost of uncertainty from 4.90\% to 1.72\%. Note that for each of the three values of $\epsilon$, we redraw the samples and obtain slightly different forecast values. As a result, the cost of uncertainty is not directly comparable between different values of $\epsilon$.

\subsection{Benders Decomposition for 241-Bus AC-DC System}
\label{sub_benders}
In the following, we demonstrate the performance of the Benders decomposition of our proposed OPF formulation for a system of two IEEE 118 bus test systems interconnected with a 5 bus multi-terminal HVDC grid. To this end, for the previously used test case in Fig.~\ref{24bus}, we replace the IEEE 24 bus systems with IEEE 118 bus systems \cite{IEEE118bus}. The converters C1 and C2 are connected to the AC buses 8 and 65 of the first 118 bus system. The converters C3 and C4 are connected to the AC buses 8 and 65 of the second 118 bus system. We place one wind farm with rated power of 300 MW at bus 5 of the first 118 bus system and a second wind farm with rated power of 600 MW at bus 64 of the second 118 bus system. The offshore wind farm with rated power of 400 MW remains at bus 3 of the HVDC grid. For the forecast data, we  select  control  zone  1,  11,  3  to  correspond  to  the  wind farms  at  bus  8,  65,  and  3,  respectively. We consider the time step 1 and one $N_s$ sample set.

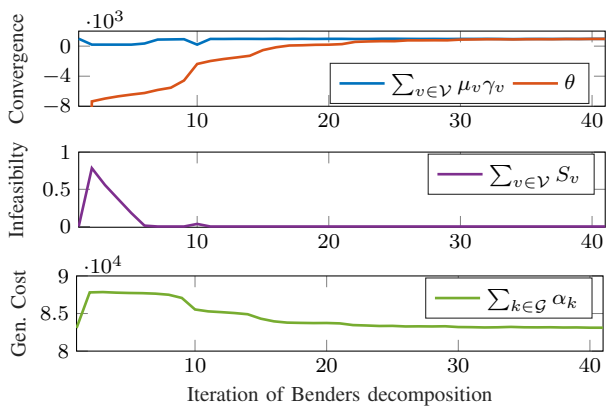
\begin{figure}
\begin{center}
		\begin{footnotesize}
%
%
\definecolor{mycolor1}{rgb}{0.00000,0.44700,0.74100}%
\definecolor{mycolor2}{rgb}{0.85000,0.32500,0.09800}%
\begin{tikzpicture}

\begin{axis}[%
width=7cm,
height=1.0cm,
scale only axis,
xmin=1,
xmax=41,
xlabel style={font=\color{white!15!black}},
ymin=-8000,
ymax=2000,
ylabel style={font=\color{white!15!black}},
ylabel={Convergence},
axis background/.style={fill=white},
scaled y ticks=base 10:-3,
ytick = {-8000,-4000,0},
legend style={legend pos = south east, legend cell align=left, align=left, draw=white!15!black,legend columns=2}
]
\addplot [color=mycolor1, line width=1pt]
  table[row sep=crcr]{%
1	983.897831292321\\
2	199.957874468446\\
3	198.933151667558\\
4	198.427469422988\\
5	198.805935521624\\
6	348.810341757368\\
7	872.455393280894\\
8	895.675817175817\\
9	918.917381873338\\
10	198.79789871575\\
11	956.630773481217\\
12	953.420987517974\\
13	966.228264615029\\
14	964.763953196196\\
15	956.612554268955\\
16	973.443569788294\\
17	955.613417451383\\
18	970.643137066701\\
19	956.456336161193\\
20	970.133614534312\\
21	969.97387081905\\
22	949.457246063676\\
23	950.241188962236\\
24	978.759650783284\\
25	980.114118982893\\
26	966.281194449521\\
27	955.287551011387\\
28	960.947592865044\\
29	949.35002217988\\
30	961.009767595504\\
31	960.14055092305\\
32	969.367885251076\\
33	956.70722857172\\
34	933.720590227443\\
35	955.121697727546\\
36	957.978367432074\\
37	953.504507207434\\
38	958.265436814492\\
39	954.059519583299\\
40	967.127307676634\\
41	967.034729661445\\
};
\addlegendentry{$\sum_{v \in \mathcal{V}} \mu_v \gamma_v $}

\addplot [color=mycolor2, line width=1pt]
  table[row sep=crcr]{%
1	-100000\\
2	-7354.19917231061\\
3	-6974.58461547197\\
4	-6678.7058806676\\
5	-6445.6020841874\\
6	-6236.88823963069\\
7	-5830.03697526907\\
8	-5531.63022446223\\
9	-4574.78197253839\\
10	-2373.64780027012\\
11	-1979.90763812003\\
12	-1717.08340656352\\
13	-1505.28732176383\\
14	-1282.01619275043\\
15	-530.261756768066\\
16	-151.182252076123\\
17	82.1045362925146\\
18	116.653178820282\\
19	177.035009980793\\
20	192.995081539693\\
21	266.917933473539\\
22	565.581144501661\\
23	623.851728968313\\
24	661.611617283323\\
25	657.91025620618\\
26	766.239843412624\\
27	767.794596317242\\
28	778.470184296494\\
29	772.273197442937\\
30	869.419262384071\\
31	889.206159738207\\
32	917.103332797013\\
33	913.991331307778\\
34	891.694145952932\\
35	917.449791587714\\
36	922.979422697348\\
37	916.15272056334\\
38	944.408364079803\\
39	939.436171696761\\
40	957.593608847707\\
41	959.902164912917\\
};
\addlegendentry{$\theta$}

\end{axis}
\end{tikzpicture}%
 \\ 
%
%
\definecolor{mycolor1}{rgb}{0.85000,0.32500,0.09800}%
\definecolor{mycolor4}{rgb}{0.49400,0.18400,0.55600}%
\definecolor{mycolor5}{rgb}{0.46600,0.67400,0.18800}%
\begin{tikzpicture}

\begin{axis}[%
width=7cm,
height=1.0cm,
scale only axis,
xmin=1,
xmax=41,
xlabel style={font=\color{white!15!black}},
ymin=-0.01,
ymax=1,
ylabel style={font=\color{white!15!black}},
ylabel={Infeasibilty},
axis background/.style={fill=white},
ytick = {0,0.5,1},
legend style={legend cell align=left, align=left, draw=white!15!black}
]
\addplot [color=mycolor4, line width=1pt]
  table[row sep=crcr]{%
1	0\\
2	0.783902249213054\\
3	0.556944662913126\\
4	0.368130451068271\\
5	0.180719572227518\\
6	0.0118755214665156\\
7	0\\
8	0\\
9	0\\
10	0.034425259107785\\
11	0\\
12	0\\
13	0\\
14	0\\
15	0\\
16	0\\
17	0\\
18	0\\
19	0\\
20	0\\
21	0\\
22	0\\
23	0\\
24	0\\
25	0\\
26	0\\
27	0\\
28	0\\
29	0\\
30	0\\
31	0\\
32	0\\
33	0\\
34	0\\
35	0\\
36	0\\
37	0\\
38	0\\
39	0\\
40	0\\
41	0\\
};
\addlegendentry{$\sum_{v \in \mathcal{V}} S_v$}
\end{axis}
\end{tikzpicture}%
 \\
%
%
\definecolor{mycolor1}{rgb}{0.00000,0.44700,0.74100}%
\definecolor{mycolor4}{rgb}{0.49400,0.18400,0.55600}%
\definecolor{mycolor5}{rgb}{0.46600,0.67400,0.18800}%
\begin{tikzpicture}

\begin{axis}[%
width=7cm,
height=1.0cm,
scale only axis,
xmin=1,
xmax=41,
xlabel style={font=\color{white!15!black}},
xlabel={Iteration of Benders decomposition},
ymin=80000,
ymax=90000,
ylabel style={font=\color{white!15!black}},
ylabel={Gen. Cost },
ytick = {80000,85000,90000},
axis background/.style={fill=white},
legend style={legend cell align=left, align=left, draw=white!15!black}
]
\addplot [color=mycolor5, line width=1pt]
  table[row sep=crcr]{%
1	83070.6840626496\\
2	87810.9954829995\\
3	87853.7655796036\\
4	87777.689426305\\
5	87732.3639516809\\
6	87708.6082599623\\
7	87643.1377072643\\
8	87500.3103286906\\
9	87072.863683774\\
10	85539.0399538755\\
11	85287.7352740987\\
12	85183.1104079665\\
13	85064.3480832965\\
14	84904.7613280696\\
15	84312.2176635315\\
16	83958.3134780533\\
17	83790.1402017263\\
18	83754.8004358245\\
19	83729.6066409409\\
20	83740.4170549498\\
21	83687.8391262332\\
22	83462.9297869663\\
23	83409.5077952911\\
24	83342.2455366713\\
25	83357.4333808704\\
26	83280.8169538562\\
27	83295.6991696108\\
28	83280.8764995741\\
29	83305.0305824736\\
30	83199.5968453252\\
31	83184.2375627324\\
32	83147.1066012414\\
33	83168.2940389903\\
34	83220.097027686\\
35	83167.915393477\\
36	83159.7615528696\\
37	83172.2669848817\\
38	83139.7777607848\\
39	83150.054698375\\
40	83115.161517288\\
41	83112.8148270385\\
};
\addlegendentry{$\sum_{k \in \mathcal{G}} \alpha_k$}
\end{axis}
\end{tikzpicture}%
 \\ 
		\end{footnotesize}
		\end{center}
		\vspace{-12pt}
	\caption{For a 241-bus system, the convergence characteristics of the proposed decomposition algorithm are displayed. The upper plot shows the sum of the objective values of the penalized subproblems and the auxiliary variable $\theta$. The middle plot shows the infeasibility and the lower plot the generation cost.}
	\label{BendersConv}
\end{figure}
We solve the decomposed OPF formulation for a uniform penalty parameter of $\mu_v = 100$. We select $\theta_{min}$ to be $-10^{8}$. As convergence criterion, we assume that the difference between the auxiliary variable and the sum of the objective value of the optimality subproblems is less than $10^{-4}$ of the overall objective value. Furthermore, the sum of the feasibility subproblems $S_v$ should be lower than $10^{-4}$. In Fig.~\ref{BendersConv} the convergence characteristics of the decomposed formulation are shown. 
In the first iterations, feasibility cuts are included, which lead to an increase in the generation cost for the forecasted system state. 
As the optimality cuts from the subproblems are gradually included the generation cost decreases. At the same time, the difference between auxiliary variable and the sum of the objective values of the subproblems decreases until the algorithm converges after 41 iterations. The solving time for one instance of the subproblem is on average 0.45 s and for the master problem on average 0.65 s. Note that we observe a decrease in the resulting numerical accuracy for the decomposed problem compared with solving the original optimization problem. If we solve the problem with a different penalty term, we can directly include the feasibility cuts from previous iterations to the master problem to speed up the convergence of the Benders algorithm.

\section{Conclusions}\label{V}
In this work, we propose a tractable formulation of a chance constrained AC-OPF for interconnected AC and HVDC grids, which uses the semidefinite relaxation of the AC-OPF and can provide guarantees regarding (near-)global optimality. We include control policies related to active power, reactive power, and voltage, in particular of HVDC converters. To enhance scalability and numerical stability, we split the semidefinite matrix in parts corresponding to each individual subsystem, i.e. different AC or DC grids. By using a penalty term on active power losses, we propose a systematic method to identify suitable penalty weights to obtain rank-1 solution matrices. To facilitate computational tractability, we  propose  a  decomposition  of  our  AC-OPF  formulation  using Benders decomposition and show its success on a 214-bus AC-DC system. For a test case of two IEEE 24 bus AC grids interconnected through an HVDC grid, using realistic forecast data, we show that a chance constrained DC-OPF leads to violations of the considered joint chance constraint whereas our proposed approach complies with all constraints. To match the empirical closely with the maximum allowable joint chance constraint violation probability, we propose a heuristic adjustment procedure for the scenario-based uncertainty set by discarding worst-case samples which allows us to reduce the cost of uncertainty. Our future work will focus on including (i)~N-1 security and post-contingency HVDC corrective control, and (ii)~successive penalization techniques.  

\section*{Acknowledgment}
The authors would like to thank Pierre Pinson for sharing the forecast data, Jalal Kazempour for his advice on Benders decomposition and the anonymous reviewers for their valuable comments.

\bibliographystyle{IEEEtran}
%
%
%
\bibliography{Bib}

\ifCLASSOPTIONcaptionsoff
\newpage
\fi

\begin{IEEEbiography}[{\includegraphics[width=1in,height=1.25in,clip,keepaspectratio]{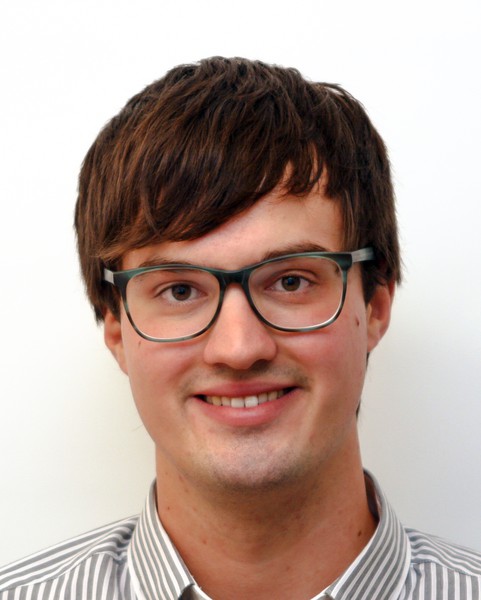}}]{Andreas Venzke} (S'16) received the M.Sc. degree in Energy Science and Technology from ETH Zurich, Zurich, Switzerland in 2017. He is currently working towards the Ph.D. degree at the Department of Electrical Engineering, Technical University of Denmark (DTU), Kongens Lyngby, Denmark. His research interests include power system operation under uncertainty and convex relaxations of optimal power flow.
\end{IEEEbiography}
\vspace{-15cm}
\begin{IEEEbiography}[{\includegraphics[width=1in,height=1.25in,clip,keepaspectratio]{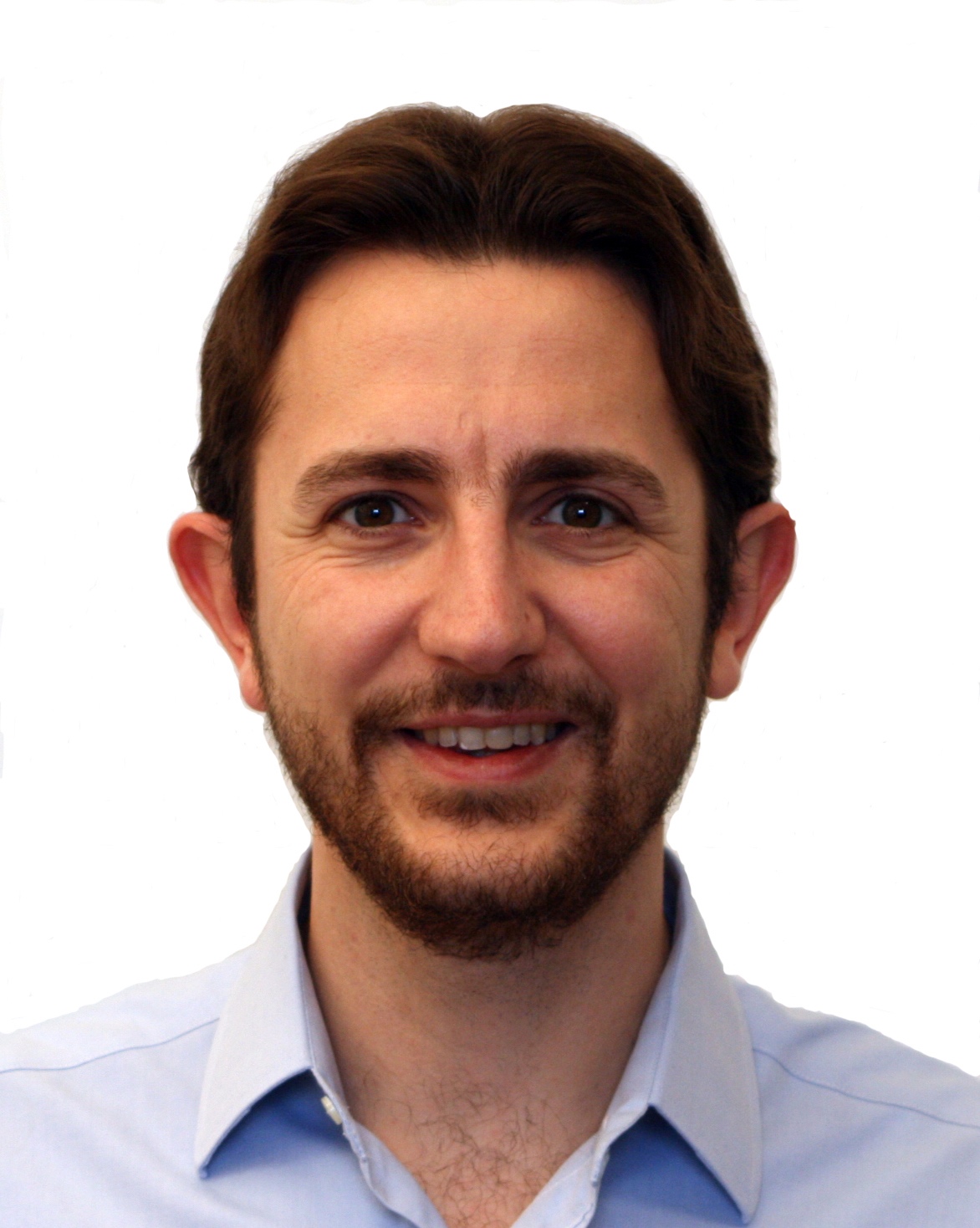}}]{Spyros Chatzivasileiadis} (S'04, M'14, SM'18) is an Associate Professor at the Technical University of Denmark (DTU). Before that he was a post-doctoral researcher at the Massachusetts Institute of Technology (MIT), USA and at Lawrence Berkeley National Laboratory, USA. Spyros holds a PhD from ETH Zurich, Switzerland (2013) and a  Diploma in Electrical and Computer Engineering from the National Technical University of Athens (NTUA), Greece (2007). In March 2016 he joined the  Center of Electric Power and Energy at DTU. He is currently working on power system optimization and control of AC and HVDC grids, including  semidefinite relaxations, distributed optimization, and data-driven stability assessment.
\end{IEEEbiography}

\end{document}